\documentclass[11pt,a4paper]{article}
\usepackage{amsthm,amscd,amssymb}
\usepackage{wrapfig}
\usepackage{epic,eepic,pstricks}
\usepackage{mathrsfs}
\usepackage{rotating}
\usepackage{placeins}
\usepackage{amsmath}
\usepackage{amsfonts}
\usepackage{graphicx}
\usepackage{psfrag}
\usepackage[font=small,skip=-10pt]{caption}

\usepackage{siunitx} 
\usepackage{pfnote} 

\newtheorem{theorem}{Theorem}
\newtheorem{corollary}{Corollary}
\newtheorem{proposition}{Proposition}
\newtheorem{lemma}{Lemma}
\def\vs{\vspace{5mm}}

\def\M{{\mathcal M}}
\def\N{{\mathcal N}}
\def\C{{\mathcal C}}
\def\Khat{\widehat{K}}
\def\phat{\widehat{p}}
\def\that{T}

\newcommand{\bm}[1]{\mbox{\boldmath $#1$}}

\newenvironment{Proof}{\noindent { \textbf{Proof.}}}{\hfill$\blacksquare$ \vs}

\newenvironment{Remark}{\stepcounter{theorem} \noindent {\textbf{Remark 
\thetheorem}.} 
}{}

\makeatletter
\let\c@proposition\c@theorem
\let\c@corollary\c@theorem
\let\c@lemma\c@theorem
\let\c@definition\c@theorem
\let\c@example\c@theorem
\makeatother

\usepackage[utf8]{inputenc}
\usepackage{verbatim}

\usepackage{varioref}
\usepackage[breaklinks=true]{hyperref}  
\usepackage{breakcites}
\usepackage{cleveref}

\usepackage{afterpage}

\usepackage{microtype} 
\usepackage{booktabs} 
\usepackage{float} 
\hypersetup{colorlinks=true,citecolor=blue,linkcolor=red,urlcolor=blue}

\usepackage{abstract} 

\newcounter{mnotecount}

\newcommand{\mnotex}[1]
{\protect{\stepcounter{mnotecount}}$^{\mbox{\footnotesize 
$\spadesuit$\themnotecount}}$ 
\marginpar{
\raggedright\tiny\em
$\!\!\!\!\!\!\,\spadesuit $\themnotecount: #1} }

\title{
McGehee regularization of general $SO(3)$-invariant potentials and applications to stationary and 
spherically symmetric spacetimes}

\author{Pablo Galindo\\[2mm] 
\small{Dept. de Geometría y Topología}\\
\small{Universidad de Granada,}\\
\small{Campus de Fuentenueva s/n,}\\
\small{18071 Granada, Spain.}\\ 
\href{mailto:pablogsal@correo.ugr.es}{{\small pablogsal@correo.ugr.es}} 
\vspace{-5mm}
\and
Marc Mars\\[2mm] 
\small{Inst. de Física Fundamental y Matemáticas}\\
\small{(IUFFyM),}\\
{\small Universidad de Salamanca,}\\
{\small Plaza de la Merced s/n 37008 Salamanca, Spain.}\\ 
\href{mailto:marc@usal.es}{{\small marc@usal.es}} 
\vspace{-5mm}
}
\date{}

\begin{document}

\maketitle 
\thispagestyle{empty}


\begin{abstract}

The McGehee regularization is a method to study the singularity
at the origin of the dynamical system describing a point particle in a plane moving under the action of a
power-law potential. It was used by Belbruno and Pretorius \cite{belbruno2011dynamical}
to perform a dynamical system regularization of
the singularity at the center of the motion of massless test
particles in the Schwarzschild spacetime. In this paper, we generalize
the McGehee transformation so that we can regularize the 
singularity at the origin of the dynamical system describing 
the motion of causal geodesics (timelike or null)
in any stationary and spherically symmetric  spacetime
of Kerr-Schild form. We first show that the geodesics for both massive and
massless particles can be described globally in the Kerr-Schild spacetime as
the motion of a Newtonian point particle in a suitable radial potential and
study the conditions under which the central singularity can be
regularized using an extension of the McGehee method. As an example,
we apply these results  to causal geodesics in the Schwarzschild and
Reissner-Nordstr\"om spacetimes. Interestingly, the geodesic
trajectories in the whole
maximal extension of both spacetimes can be described by a single
two-dimensional phase space with non-trivial topology. This topology
arises from the presence
of excluded regions in the phase space
determined by the condition that the tangent vector of the geodesic be
causal and future directed.


\end{abstract}


\section{Introduction} Kerr-Schild metrics \cite{kerr1965new} are a
well-known Ansatz to solve the Einstein field equations and leads to
many physically important exact solutions of the four-dimensional
case, such as the Schwarzschild black hole, the Reissner-Nordstr\"om,
the Kerr black hole, the charged Kerr–New\-man black hole, the Vaidya
radiating star, Kinnersley photon rocket, pp-waves and also some of
their higher dimensional analogues \cite{malek2012exact}. Kerr-Schild
metrics have played a crucial role in the discovery of rotating black
holes in higher dimensions \cite{myers1986black,gibbons2005general} as
well as in the recent work on so-called higher order
gravities \cite{anabalon2009kerr,anabalon2011remarks}. Also, most
static and spherically symmetric spacetimes can be displayed in
Kerr-Schild form and their analysis conforms a field that continues
giving interesting results nowadays
\cite{parry2012survey,hackmann2008analytic}. Two of the best known
static and spherically symmetric metrics which have been studied
extensively and remain an area of current research are the
Schwarzschild metric and the Reissner-Nordstr\"om metric. Although the
behavior of the geodesics in both metrics is well-known,
the geodesic equations have a large number of dynamic properties that are
still providing new results, such as the characterization of the circular motion in the Reissner-Nordstr\"om spacetime
for neutral and charged particles \cite{pugliese2011circular,pugliese2011motion} or the dynamics of the chaotic motion in the Schwarzschild black hole surrounded by an external halo \cite{de2000chaos}. The dynamical system approach to the
analysis of the geodesic flow in these spacetimes and their rotating Kerr generalizations
is a  novel approach which, besides
providing many new and interesting results, also describes known results from a different perspective. Examples
are the homoclinic orbits that asymptotically approach the unstable branch
of circular orbits \cite{levin2008periodic,levin2009homoclinic,perez2009homoclinic,misra2010rational} or
the fact that perturbation of the geode\-sic flow possesses a chaotic
invariant set \cite{moeckel1992nonintegrable,levin2000gravity,suzuki1999signature}. One
of the advantages of this method is that a great amount
of information can be obtained without integrating the geodesic equations.
Also, by the use
of ``blow-up'' techniques in the dynamical system we can describe the
behavior of the geodesic equations near the singularity, of which
little is known. One of the most recent works along this line
\cite{belbruno2011dynamical} has analyzed the null case of the
geodesic flow by the use of the McGehee regularization
\cite{mcgehee1981double}, which is a method designed to deal with the
singularities at the center in the motion of Newtonian particles subject to a central power-law potential. In the context of geodesics in Schwarzschild, the
limitation of the MacGehee  method is the restriction to
power-law potentials, which prevents its application to 
timelike geodesics. This is one of the reasons why null geodesics only
where treated in \cite{belbruno2011dynamical}. Also, the standard
McGehee method involves a somewhat complicated phase space which obscures
the analysis. Other approaches to this problem
\cite{stoica1997schwarzschild} 
have studied timelike geodesic in
Schwarzschild by the
use of a variation of the McGehee method. However, the approach is such 
that one deals with  a one-parameter
family of energy-dependent dynamical systems in which only one curve
in each phase space is relevant. This obviously 
obscures and complicates unnecessarily  the results (in fact, this
drawback was not explicitly noticed in \cite{stoica1997schwarzschild}).

In this paper we generalize the McGehee regularization so that
we can deal with central potentials of a very general  form. With this
method we can treat not only general causal geodesics in Schwarzschild
but also geodesics in the Reissner-Nordstr\"om spacetime. In
fact, for a substantial fraction of the paper we work in full
generality in stationary and spherically symmetric spacetimes of
Kerr-Schild form, of which the previous are just particular
cases. There are several possible approaches to derive the geodesics
equations in such spacetimes.  Explicit computation of the Christoffel
symbols is tedious and not particularly enlightening. It
is particularly cumbersome to incorporate the conserved quantities associated to
Killing vectors into the equations.  
More straightforward
and convenient
is the use of Hamiltonian methods which, in particular, allows
for the incorporation of conserved quantities into the system in a
straightforward way. 
Once we have the geodesic equations for such spacetimes, we can
apply the generalized McGehee regularization and
subsequently analyze the phase space defined by the geodesic
equations, with particular  emphasis at the vicinity of the
singularity, where new and interesting dynamics appears. A remarkable 
fact is that the dynamics in the entire
maximal extension of the spacetime can be described 
in a single two-dimensional phase space, which has
particular importance in the Reissner-Nordstrom case. The key for
this lies in the presence of excluded regions in the phase
space arising from the condition that the trajectories correspond to
future directed causal geodesics.

The paper is organized as follows: In \cref{stationary} we follow
a simple way to obtain the geodesic equations for a general
stationary Kerr-Schild metric and obtain a simplified Hamiltonian
with the Killing conserved quantity already incorporated. In
\cref{sphericalkerrschild} we particularize 
to the case of stationary and spherically symmetric Kerr-Schild spacetimes.
In particular,
we find that the geodesics can be described
by a classical Hamiltonian of the form $H = T + V$ with $V$
a spherically symmetric potential. It is not at all clear a priori that
this should be possible in the whole Kerr-Schild domain. The Hamilton equations
already incorporate all the constants of motion associated to the
symmetries. We analyze under which conditions a Hamiltonian
trajectory corresponds to a causal, future directed geodesic 
of the spacetime. These conditions will be
translated into excluded regions in the corresponding phase spaces. 
In Section \ref{McGeheeregularization}
we generalize the McGehee transformation to radial potentials of very
general form and provide a method to choose the appropriate
parameter to perform the regularization. This discussion also helps clarifying the original regularization procedure proposed
by McGehee. The physical meaning of the generalized McGehee 
variables is also discussed.
In Section \ref{SW} we particularize the previous general results to
the Schwarzschild spacetime paying particular attention to the
collision manifold and to the excluded region for future-oriented
geodesics. We recover the known results on null geodesics  near
the singularity obtained in \cite{belbruno2011dynamical} 
and extend them to timelike geodesics (in fact, all causal geodesics are treated
simultaneously). As already mentioned, the understanding of the excluded
regions is crucial to have a phase space
of physical trajectories with a non-trivial topology capable of dealing 
with all Kerr-Schild patches of the Kruskal spacetime.
Finally, in Section \ref{RN} we perform a similar analysis 
for the maximal extension of the Reissner-Nordstr\"om spacetime. 

\section{Geodesic equations for a general stationary Kerr-Schild metric}
\label{stationary}

Throughout this paper, we will consider spacetimes
$\{ \M = \mathbb{R} \times (\mathbb{R}^3 \setminus \C),g \}$ where
$\C \subset \mathbb{R}^3$ is a closed subset such that
$\M$ is connected and $g$ is a Lorentzian metric of Kerr-Schild form 
\cite{kerr1965new}.
More specifically, let  $\{ x^{\alpha} \} = \{ \that, x^i \}$ ($\alpha,\beta,
\cdots = 0,1,2,4$ and  $i,j, \cdots =1,2,3$)  be Cartesian coordinates on 
$\mathbb{R} \times \mathbb{R}^3$ and endow
$\M$ with the Minkowski metric
$\eta = - d \that^2 + \delta_{ij} dx^i dx^j $. Let $\bm{K}$ be a smooth
one-form on $\M$ which is null with respect to the metric $\eta$ and
$h: \M \longrightarrow \mathbb{R}$ a smooth function.
The metric $g$ being of Kerr-Schild form means that it takes the form
\begin{align}
g_{\alpha\beta}  &= \eta_{\alpha\beta}  +h K_{\alpha} K_{\beta}. 
\label{eq:KerrSchildformmetric} 
\end{align} 
It is well-known (and immediate to check) that
 the inverse metric $g^{-1}$ is
\begin{gather*}
(g^{-1})^{\alpha \beta}=\eta^{\alpha \beta} - h K^\alpha K^\beta,
\end{gather*}
where all Greek indices are raised and lowered with the Minkowski metric
$\eta$.
This expression shows, in particular, that the one-form $\bm{K}$ is also
null in the metric $g$. 
We will assume from now on that neither $\bm{K}$ nor
$h$ vanish on a non-empty open set on $\M$.

Our aim in this section is to study the geodesic equations for a
Kerr-Schild metric assuming the spacetime to be stationary
with Killing vector $\xi = \partial_{\that}$.
It is clear from (\ref{eq:KerrSchildformmetric})
that $\xi$ is a Killing vector of $g$
if and only 
\begin{gather}
(\pounds_{\xi} h ) \bm{K} \otimes \bm{K} 
+ h (\pounds_{\xi}  \bm{K} ) \otimes \bm{K} + 
+ h \bm{K}  \otimes (\pounds_{\xi}  \bm{K} ) =0.
\label{Stat}
\end{gather}
where $\pounds$ denotes Lie derivative.
At any point $p \in \M$ where $\bm{K} |_p \neq 0$,
let $V_p \in T^{\star}_p \M$ be a vector subspace such that
$T^{\star}_p M = \langle K|_p
\rangle \oplus V_p$ and use this direct sum to decompose
$\pounds_{\xi} {\bm K} |_p = C|_p \bm{K} + 
\bm{U}|_p$.
Inserting this into (\ref{Stat})
yields
\begin{gather*}
\left ( \pounds_{\xi} h + 2 C h \right ) 
\bm{K} \otimes {\bm K}
+ h \left ( \bm{U} \otimes  {\bm K} + \bm{K} \otimes \bm{U} \right ) |_p =0
\end{gather*}
which is equivalent to $h {\bm U} |_p =0$ and
$ (\pounds_{\xi} h + 2 C h) |_p =0$. Using the fact that neither $h$ nor
$\bm{K}$ vanish on a non-trivial open set, it follows that 
$\xi =  \partial_\that$ is a Killing vector of $g$
if and only if there exists a smooth function $C: \M \longrightarrow \mathbb{R}$
such that 
$\pounds_{\xi} \bm{K} = C \bm{K}$ and
$\pounds_{\xi} h = - 2 C h$. Let $f_0 : \mathbb{R}^3 \setminus \C 
\longrightarrow \mathbb{R}$ be any smooth positive function and
let $f : \mathbb{R} \times ( \mathbb{R}^3  \setminus \C )
\longrightarrow \mathbb{R}$ be the unique solution of $\partial_{\that} f
 + C f = 0$ with initial data $f |_{\that =0} = f_0$. It is immediate
to check that $f > 0$ everywhere. Defining $h' := \frac{h}{f^2}$
and $\bm{K}^{\prime} := f \bm{K} $, they satisfy
\begin{align*}
\pounds_{\xi} h^{\prime} = 0, \\
\pounds_{\xi} {\bm K}^{\prime} = 0,
\end{align*}
while the metric $g$ takes the form
\begin{align}
g_{\alpha\beta}  &= \eta_{\alpha\beta}  +h^{\prime} K^{\prime}_{\alpha} 
K^{\prime}_{\beta}.
\label{metricprime} 
\end{align}
Dropping the primes, it follows that $\xi$ is a Killing vector for
$g$ if and only $h$ and $\bm{K}$ can be selected to be Lie constant
along $\xi$. We assume this from now on.

In the Minkowskian coordinates $\{ x^{\alpha} \}$ 
let us write $K^\alpha=( \Khat, \vec{K})$ where $\Khat$ satisfies
$\Khat^2 = \vec{K}^{\, 2} := K^i K_i$ and Latin indices are raised and
lowered
with the Euclidean metric $\delta_{ij}$. The Killing vector $\xi$ is timelike
on the set $\{ p \in \M ; \, \, h \vec{K}^{\,2} |_p   < 1\}$, null on the set
$\{ p \in \M; \, \, h \vec{K}^{\, 2} |_p = 1\} $ and 
spacelike on the set  $\{ p \in \M ; \, \, h \vec{K}^{\, 2} |_p   > 1\}$. Note 
also
that  we are not assuming $\bm{K}$ to be future directed or
past directed everywhere, so
that a priori $\Khat$ may change sign. 

In any spacetime $(\M,g)$, affinely parametrized geodesics
are the solutions of the Hamilton equations of the Hamiltonian
\begin{equation}
H=\frac{1}{2}(g^{-1})^{\alpha \beta} p_\alpha p_\beta
\end{equation}
defined on the cotangent bundle of $\M$. 
The Hamilton
equations fix $\bm{p}= g (u, \cdot)$
where $u$ is 
the tangent vector to the geodesic.
Using  the explicit expression (\ref{eq:KerrSchildformmetric})
for the metric, this Hamiltonian takes the form
\begin{equation}
\label{Hamiltonkerrschild} 
H= \frac{1}{2} \left(
\eta^{\alpha \beta} p_\alpha p_\beta -  h (K^\alpha p_\alpha)^2
\right).
\end{equation}
Given that $\xi$ is a Killing vector,
the quantity $E := - \bm{p} (\xi)$ is conserved along geodesics.
Note also that, with this definition,
\begin{equation}
\label{Kdotp}
K^{\alpha} p_{\alpha} = - E \Khat + \vec{K} \cdot \vec{p},
\end{equation}
where we have written $\bm{p} = \{ \phat,
\vec{p} \, \}$ and dot means scalar product with $\delta_{ij}$.

The Hamiltonian itself is a conserved quantity with the value of
$H=-\frac{1}{2} \mu$ where $\mu= {0,\pm 1}$ depending on whether the
geodesic is timelike ($\mu=1$), spacelike ($\mu=-1$) or null
($\mu=0$). Inserting (\ref{Kdotp}) and the conserved quantity $E$
into (\ref{Hamiltonkerrschild})
the following Hamiltonian arises naturally
\begin{equation}
\label{KerrSchildHamilton}
H^{\prime} := H + \frac{1}{2} E^2 = 
\frac{1}{2} \left( \vec{p}^{\,2}- h \left( \vec{K} \cdot \vec{p} - E 
\Khat \right)^2 \right),
\end{equation} 
which is now defined on the cotangent 
bundle of $\mathbb{R}^{3} \setminus \C$. 

The interest of this Hamiltonian lies in the fact (easy to check)
that if a curve $(\that (s), \vec{x}(s) \}$ is a geodesic
in $(\M,g)$ with tangent vector $u$ satisfying $g (u,u) = -
\mu$ and conserved quantity $\bm{p} (u ) = -E$,
then $\{ \vec{x}(s) \}$ is the projection to the base
space $\mathbb{R}^3 \setminus \C$ of a solution of the Hamilton equations 
of (\ref{KerrSchildHamilton}) satisfying
\begin{equation}\label{HamiltonianEnergyMu}
H' = \epsilon := \frac{1}{2} \left ( E^2 - \mu \right )
\end{equation}
along the curve and $\that(s)$ satisfies the ODE
\begin{equation}
\left (  1 -  h \vec{K}^{\, 2}  \right ) \frac{d \that}{ds}
+ h \Khat \vec{K} \cdot \frac{d \vec{x}}{ds} = E,
\label{eq:conservedenergy}
\end{equation}
which is simply the explicit form for $g(u, \xi) = -E$ in the
Cartesian coordinates $\{ \that, \vec{x} \}$. The converse to this
statement will be addressed in the following section in the spherically
symmetric case.

\section{Geodesic equations for a stationary and spherically symmetric Kerr-Schild metric}\label{sphericalkerrschild}

In this section we want to particularize the problem to the spherically 
symmetric setting. So, we assume
the group of rotations $SO(3)$ acting on $\M$ as 
\begin{align*}
SO(3) \times \M & \longrightarrow \M, \\
(R , (T,\vec{x})) & \longrightarrow (T, R(\vec{x}))
\end{align*}
to be an isometry of $g$. Note that, for this definition to make sense, the set 
${\cal C}$ must be invariant
under the SO(3) action, which we assume from now on. The isometry condition 
requires $\pounds_{\vec{\zeta}} \, (h \bm{K} \otimes
\bm{K} ) = 0$, for any generator $\zeta$ of the group $SO(3)$. Pulling back this
relation to the orbits of the isometry group and using the fact that the only 
symmetric 2-covariant tensor on the sphere
which is  invariant
under $SO(3)$ is a constant times the standard metric on the sphere, it follows 
that $h \bm{K} \otimes \bm{K}$ pulls back to zero
on the $SO(3)$ orbits. Since $h$ does not vanish on open sets, we conclude 
that $\bm{K}$ itself
pulls back to zero on these surfaces. Given the stationarity condition, 
this is equivalent to the existence of a smooth function $f:  \mathbb{R}^3 
\setminus {\cal C} \longrightarrow \mathbb{R}$ such that
$\vec{K} = f \frac{\vec{x}}{|\vec{x}|}$ where
$|\vec{x}|: = \sqrt{\vec{x} \cdot \vec{x}}$. Hence,
\begin{align}
K^\alpha=\left( \hat{K}, \vec{K} \right)=\left( \hat{K}, 
f \frac{\vec{x}}{|\vec{x}|} \right), \quad \quad
\mbox{with} \quad  \hat{K}^2 =f^2. 
\label{Kalpha}
\end{align}
The following lemma gives the most general form of $g$ under a mild additional 
restriction.
\begin{lemma}
Assume that $h$ is non-zero on a dense set, that the null vector $K^\alpha$ 
does not have any flat zero (i.e. a point where 
$K^{\alpha}$ and all its derivatives vanish)  and that $(\M,g)$ is stationary 
and spherically symmetric.  Then 
$h$ and $K^\alpha$ can be chosen so that $h(\vec{x})$ is spherically symmetric 
and
\begin{equation*}
K^\alpha=\left( \sigma, \frac{\vec{x}}{|\vec{x}|} \right), 
\end{equation*}
where $\sigma = \pm 1$ is a constant on $\M$.
\end{lemma}
\begin{Proof}
The condition that $f$ has no flat zeros implies that $f$ (and hence
$\bm{K}$) cannot vanish on a 
non-empty open set. So, as discussed
in Section \ref{stationary}, we can assume $\pounds_{\xi} h =0$ and $\pounds_{\xi} 
\bm{K} =0$ where $\xi = \partial_T$, and that (\ref{Kalpha}) holds.
Let ${\cal S}_{I}$ be the collection of arc-connected components of $\{ f \neq 
0 \} \subset \mathbb{R}^3 \setminus {\cal C}$. 
On each one of these open sets we have $\hat{K} = \sigma_I f$, where $\sigma_I 
= \pm 1$ is constant on
${\cal S}_I$. Let ${\cal S}_+$ be the union of components ${\cal S}_I$ with 
$\sigma_I = +1$
and ${\cal S}_-$ be the union of components ${\cal S}_I$ with $\sigma_I = - 1$ 
and assume that both are non-empty. Since ${\cal S}_+ 
\cup {\cal S}_{-}$ is dense in $\mathbb{R}^3 \setminus {\cal C}$ and the latter 
is connected
it follows that there exists a point
$p \in \mathbb{R}^3 \setminus {\cal C}$ that can can be approached by a 
sequence $\{ p^+_i \in {\cal S}_+ \}$
and by a sequence  $\{ p^{-}_i \in {\cal S}_{-} \}$. Since $\hat{K}$ is smooth 
everywhere, in particular at $p$, it follows that necessarily
$f$ and all its derivatives vanish at $p$, against assumption. Thus, either 
${\cal S}_- = \emptyset$ (and we can write
$\hat{K} = f$ everywhere) or 
${\cal S}_{+} = \emptyset$ (and we can write
$\hat{K} = - f$ everywhere). Consequently, the Kerr-Schild metric takes the 
form $g = \eta + h f^2 \bm{K}' \otimes \bm{K}'$ with 
$K'{}^{\alpha} = ( \sigma, \frac{\vec{x}}{|\vec{x}|} )$. Defining $h' = h f^2$, 
and given the spherically symmetric invariance
of $K'{}^{\alpha}$, it follows immediately that $g$ is spherically symmetric if 
and only if $h'$ is spherically symmetric. Dropping the 
primes in $K'^{\alpha}$ and $h'$ the lemma follows. 
\end{Proof}

\begin{Remark}\label{completingeqsremark}
As a consequence of this lemma,  the Hamiltonian $H'$ in 
\cref{KerrSchildHamilton} 
takes the form
\begin{equation}\label{eq:sphericalHamiltonian}
H' = \frac{1}{2} \vec{p}^{\,2}- \frac{h}{2} \left( \frac{\vec{x} \cdot 
\vec{p}}{|\vec{x}|} 
- \sigma  E  \right)^2.
\end{equation}
An important question is to what
extent the field
equations of this Hamiltonian 
reproduce the information concerning the
geodesics of $g$. Note first that the Hamilton equation 
$\dot{\vec{x}} = \frac{\partial H'}{\partial \vec{p}}$ reads
explicitly
\begin{equation}
\dot{\vec{x}}  = \vec{p} - \left . h(\vec{x}) \left ( \frac{
\vec{x} \cdot \vec{p}}{|x|} - \sigma E \right ) \frac{\vec{x}}{|\vec{x}|}
\right .
\label{vel2_a}
\end{equation}
which can be written in matrix form as  (we denote
by $\vec{x}^{\, T}$  the transpose of the vector column $\vec{x}$
and by $\mbox{Id}$ the identity matrix)
\begin{gather}
\dot{\vec{x}}= \left (
\mbox{Id} - h \frac{ \vec{x} \vec{x}^{\, T} }{|\vec{x}|^2}
\right ) \vec{p} + \sigma E h \frac{\vec{x}}{|\vec{x}|}.
\label{InitialDataRestrictions}
\end{gather}
Now, the relationship between the four-velocity $u^{\alpha}$
of a geodesic and the corresponding four-momentum $p_{\alpha} = 
g_{\alpha\beta} u^{\beta}$ is obviously invertible. For a geodesic
$\{T(s),\vec{x}(s)\}$, the four-velocity is $u = \dot{T}(s) \partial_T
+ \dot{\vec{x}}(s) \partial_{\vec{x}}$. Lowering indices and using
$K_{\alpha} = - \sigma dT  +\frac{\vec{x}}{|\vec{x}|} d \vec{x}$ it follows
\begin{gather*}
\bm{p} = \left ( (h-1) \dot{T} - h \sigma \frac{\vec{x}
\cdot \dot{\vec{x}}}{|\vec{x}|}
\right ) dT + \left ( \dot{\vec{x}} + h \left  ( 
\frac{\vec{x} \cdot \dot{\vec{x}}}{|\vec{x}|} - \sigma \dot{T} 
\right ) \frac{\vec{x}}{|\vec{x}|} \right ) d \vec{x}.
\end{gather*}
Since $E = - \bm{p} (\partial_T)$, we also have $\bm{p} = - E dT
+ \vec{p} \cdot d \vec{x}$ or, equivalently,
\begin{gather}
E = (1 - h) \dot{T} + h \sigma \frac{\vec{x} \cdot \dot{\vec{x}}}{|\vec{x}|},
\label{E}\\
\vec{p} = \dot{\vec{x}} + h \left ( \frac{\vec{x} \cdot 
\dot{\vec{x}}}{|\vec{x}|} - \sigma \dot{T} \right )
\frac{\vec{x}}{|\vec{x}|}. \label{p}
\end{gather}
The first equation is exactly equation (\ref{eq:conservedenergy})
for the case under consideration and must be added
to the Hamiltonian system (\ref{eq:sphericalHamiltonian})
in order to describe the geodesics.
Concerning the second equation, its relationship to
equation (\ref{vel2_a}) is as follows. First of all,
it is immediate to check that any trajectory
satisfying (\ref{E})-(\ref{p}) also satisfies the pair of equations
(\ref{E})-(\ref{vel2_a}). To analyze 
the converse, observe
that the matrix in parenthesis in (\ref{vel2_a})
is invertible for all $h \neq 1$. So, given
$\vec{x}(s)$, this equation can be solved uniquely to obtain
$\vec{p}(s)$ and hence, assuming that (\ref{E}) holds, this solution
must be necessarily (\ref{p}). This shows the equivalence
between (\ref{E})-(\ref{p}) and (\ref{E})-(\ref{vel2_a})
at points where $h \neq 1$.  However,
at points where $h=1$ (corresponding to the
set where the Killing vector $\partial_T$ is null)
the matrix in parenthesis is the projector
orthogonal to $\vec{x}$ and hence not invertible. Thus, the component of
$\vec{p}$ parallel to $\vec{x}$ is not determined by
(\ref{InitialDataRestrictions}). It follows that, 
at points where $h=1$, the set of Hamilton equations
of $H'$ and the ODE (\ref{E}) must be supplemented by
the component of $\vec{p}$ in (\ref{p}) parallel to $\vec{x}$ which is,
for any value of $h$,
\begin{gather}
\vec{x} \cdot \vec{p} =   (1+h) (\vec{x} \cdot \dot{\vec{x}}) 
- \sigma h |\vec{x}| \dot{T}.
\label{additional}
\end{gather}
Note finally that, at points where $h=1$ the dependence
of $\dot{T}$ drops completely from (\ref{E}). Given a
solution $\{\vec{x}(s),\vec{p}(s)\}$ of the Hamiltonian
equations of $H'$, it is precisely (\ref{additional}) that allows one to solve 
for $\dot{T}(s)$ at points
satisfying $h=1$, and hence must be added to the system.
\end{Remark}

The next lemma shows that the trajectories of the Hamiltonian 
(\ref{eq:sphericalHamiltonian}) can be also
obtained by solving a much simpler Hamiltonian.

\begin{lemma}\label{Hamiltonianequivalence} 
Let $\Omega \subset \mathbb{R}^3$ be a domain and
$\{ \vec{x} \}$ Cartesian coordinates on $\Omega$.
Consider the phase space ${\cal F} := \Omega \times \mathbb{R}^3$
with global canonical coordinates $\{\vec{x}, \vec{p} \, \}$ 
and let $\pi : \Omega \times \mathbb{R}^3 \longrightarrow \Omega$
be the projection. Define on ${\cal F}$  the two Hamiltonians
\begin{align}
H' &= \frac{1}{2} \vec{p}\, {}^2- \frac{h(\vec{x})}{2}  \left( \frac{\vec{x}
\cdot \vec{p}}{|\vec{x}|}  - \sigma E  \right)^2  ,  & E \in \mathbb{R} 
\nonumber \\
\hat{H}& =\frac{1}{2} \hat{p}\, {}^2-\frac{h(\vec{x})}{2} 
\left(\frac{L^2}{|\vec{x}|^2}+ \mu \right), & L, \mu \in \mathbb{R}
\label{HamilhatH}
\end{align}
where $h : \Omega \rightarrow \mathbb{R}$ is rotationally symmetric
and $\sigma = \pm 1$. 
Denote by $\gamma(\vec{x}_0,\vec{p}_0)(s)$ (resp.
$\hat{\gamma}(\hat{x}_0,\hat{p}_0)(s)$) the $H$-trajectory (resp. 
$H'$-trajectory) passing at $s=0$ through the point $(\vec{x}_0,\vec{p}_0)$ 
(resp. $(\hat{x}_0, \hat{p}_0)$). Assume that in some
neighborhood of $\vec{x}_0$, $h(\vec{x})$ is not of the form
$h(\vec{x}) = 
\alpha |\vec{x}|^2 \left ( \beta + \gamma |\vec{x}|)^2 \right )^{-1}$
with $\alpha,\beta, \gamma \in \mathbb{R}$.
Then, the two projection curves $\pi (\gamma(\vec{x}_0,\vec{p}_0 )(s))$ and
$\pi (\hat{\gamma}(\hat{x}_0,\hat{p}_0 )(s))$ are the same if and only 
if
\begin{equation*} 
\hat{x}_0= \vec{x}_0, \quad \quad
\hspace{-1pt}  \hat{p}_0 =  \vec{p}_0 - h(\vec{x}_0) \left ( \frac{
\vec{x}_0 \cdot \vec{p}_0}{|\vec{x}_0|} - \sigma E \right ) 
\frac{\vec{x}_0}{|\vec{x}_0|},  \quad \quad
\hspace{-6pt} | \vec{x}_0 \times \vec{p}_0 |  =  |L|, \quad \quad
\hspace{-1pt} H'(\vec{x}_0, \vec{p}_0)  = \frac{1}{2} \left ( E^2 - \mu \right 
).
\end{equation*} 
\end{lemma}

\begin{Proof}
First of all, we note that a curve $\vec{x}(s)$
in $\mathbb{R}^3$  satisfying 
\begin{align}
\vec{x}(s) \times \dot{\vec{x}}(s) & = \vec{J}  \\
\frac{\dot{\vec{x}}(s)^2}{2} + V(|\vec{x}(s)|)
& =  \epsilon, 
\end{align}
where $\vec{J}$ and $\epsilon$ are constants, is
uniquely determined by the initial data $\vec{x}(0)$ and $\dot{\vec{x}}(0)$.
This is a known result of central forces in $\mathbb{R}^3$. 

Let $\{ \vec{x}(s), \vec{p}(s) \} = \gamma(\vec{x}_0,\vec{p}_0)(s)$ 
and  $\{ \hat{x}(s), \hat{p}(s) \} = \hat{\gamma}(\hat{x}_0,\hat{p}_0)(s)$.
Both Hamiltonians $H'$ and $\hat{H}$ are spherically symmetric and time
independent, so there exist constants $\vec{J}$, $\hat{J}$, $H'_0$ 
and $\hat{H}_0$  such that
\begin{gather*}
\vec{x}(s) \times \vec{p}(s) = \vec{J}, \quad \quad H'(\vec{x}(s),\vec{p}(s))
= H'_0, \\
\hat{x}(s) \times \hat{p}(s) = \hat{J}, \quad \quad \hat{H}(\hat{x}(s),
\hat{p}(s)) = \hat{H}_0.
\end{gather*}
The respective Hamilton equations imply
\begin{align}
\dot{\hat{x}}(s) & = \hat{p}(s), \label{vel1} \\
\dot{\vec{x}}(s) & = \vec{p}(s) - \left . h(\vec{x}\,) \left ( \frac{
\vec{x} \cdot \vec{p}}{|x|} - \sigma E \right ) \frac{\vec{x}}{|\vec{x}|}
\right |_{\vec{x}= \vec{x}(s), \vec{p} = \vec{p}(s)} \label{vel2}
\end{align}
and hence
\begin{gather*}
  \hat{x}(s) \times \dot{\hat{x}}(s) = \hat{J}, \quad \quad
\vec{x}(s) \times \dot{\vec{x}}(s) = \vec{J}.
\end{gather*}
We next write down explicitly $H'(\vec{x}(s),\vec{p}(s)) - H'_0 =0$.
For any vector $\vec{a}$, we can compute its square norm as
\begin{gather}
\vec{a}^2 = \frac{(\vec{x}\times \vec{a}\,)^2 + (\vec{x} \cdot \vec{a}\,)^2}{
|\vec{x}|^2}. \label{square}
\end{gather}
From (\ref{vel2}) we have
\begin{gather}
\vec{x}(s) \cdot \dot{\vec{x}}(s) = \left ( \left . \frac{}{} (\vec{x} \cdot 
\vec{p}\,)
(1-h) + \sigma h  |\vec{x}| E \right ) \right
|_{\vec{x}= \vec{x}(s), \vec{p}= \vec{p}(s)}.
\label{product1}
\end{gather}
Decomposing $\vec{p}(s)^2$ and $\dot{\vec{x}}(s)^2$
according to (\ref{square})  and 
inserting (\ref{product1}), a straightforward calculation 
transforms $(1-h) \left( \, H'_0 - H'(\vec{x}(s),\vec{p}(s) ) \, \right) =0$
into 
\begin{equation}
H'_0 = 
\frac{1}{2} \dot{\vec{x}}(s)^2 - \frac{h(\vec{x}(s))}{2} \left ( 
\frac{\vec{J}^{\,2}}{|\vec{x}(s)|^2} + E^2 - 2 H'_0 \right ) := 
\frac{\dot{\vec{x}}(s)^2}{2} 
+ V(|\vec{x}(s)|).
\label{firstpotential}
\end{equation}
where the second equality defines $V(|\vec{x}|)$.
For the trajectory $\hat{x}(s)$, the form of the Hamiltonian
$\hat{H}$  immediately implies
\begin{equation}
\hat{H}_0 = \frac{1}{2} \dot{\hat{x}}(s)^2 - \frac{h(\hat{x}(s))}{2} \left ( 
\frac{\vec{L}^2}{|\hat{x}(s)|^2} + \mu \right ) : = 
\frac{\dot{\hat{x}}(s)^2}{2} + \hat{V} (|\hat{x}(s)|),
\label{secondpotential}
\end{equation}
where the second equality defines $\hat{V}(|\hat{x}|)$.
Comparing (\ref{firstpotential}) and 
(\ref{secondpotential}) we conclude
that the two trajectories $\vec{x}(s)$ and $\hat{x}(s)$ agree
if and only if they have initial position, initial velocity 
and the respective potential functions
$V(|\vec{x}|)$ and $\hat{V}(|\vec{x}|)$ agree up to an additive
constant $c$. The condition $V(|\vec{x}|)- \hat{V}(|\vec{x}|) - c =0$ 
reads explicitly
\begin{gather*}
h(\vec{x}\,) \left ( \vec{J}^{\,2} - L^2 + \left ( E^2 - 2 H_0'-
\mu \right ) | \vec{x}|^2 \right ) = - 2 c|\vec{x}|^2.
\end{gather*}
Since by hypothesis $h(\vec{x}\,)$ is not of the form
$h(\vec{x}\,) = \alpha |\vec{x}|^2 \left ( \beta + \gamma
|\vec{x}|^2 \right )^{-1}$ in any neighborhood of $\vec{x}_0$, this
equation has as only solution $c=0$, $\vec{J}^{\,2} = L^2$ and
$H'_0 = \frac{1}{2} (E^2 - \mu ) $.  
We conclude that the trajectories
$\vec{x}(s)$ and $\hat{x}(s)$ agree if and only if
$\vec{x}_0 = \hat{x}_0$, 
$\dot{\vec{x}}|_{s=0} = \dot{\hat{x}}|_{s=0}$,
$|\vec{J}|= |L|$ and $H'_0 = \frac{1}{2} (E^2 - \mu ) $.  
Given the relation (\ref{vel2}) between velocity and momentum, the
lemma follows.
\end{Proof}

\begin{Remark}\label{Remarksigma}
It is interesting that the  Hamiltonian 
$\hat{H}$ is independent of $\sigma$, so that we will be able
to describe the geodesics in 
$(\M,g)$
both for the case when $\bm{K}$ is future directed 
(plus sign) or past directed (negative sign). Moreover, the Hamiltonian
$\hat{H}$ is a standard Hamiltonian in Newtonian mechanics for a point particle 
in
a central potential. This a substantial simplification over the original
problem of solving the geodesic equations
in a stationary and spherically symmetric spacetime of Kerr-Schild form, 
because we can exploit all the information known for trajectories of point
particles in Newtonian mechanics under the influence of a 
radial potential of the form
\begin{gather}
V(|\vec{x}|)=-\frac{h(\vec{x})}{2} \left(\frac{L^2}{|\vec{x}|^2} +  \mu \right). 
\label{potential}
\end{gather}
The main consequence of Lemma \ref{Hamiltonianequivalence} is, thus, that 
the spatial part of all geodesics in a
stationary and spherically symmetric spacetimes of Kerr-Schild form turns out to
be equivalent to the (much simpler) problem of solving the trajectory of a 
Newtonian
point particle in the potential (\ref{potential}). 
Once the spatial part of the geodesics is solved, the temporal part is dealt 
with
by solving equation (\ref{E}) (at points where $h \neq 1$)
and equation (\ref{additional}) (at points where $h=1$).
Since we are interested in causal and 
future
directed geodesics we need to find the restrictions on the initial data which 
guarantee this.
The following Proposition summarizes the results above and addresses the issue 
of future
directed initial data for both choices of $\sigma$.
\end{Remark}

\begin{proposition}\label{geodesicequationslema}
Let $\M = \mathbb{R} \times (\mathbb{R}^3 \setminus {\cal C})$ be connected 
with ${\cal C} \subset \mathbb{R}^3$ closed.
The most general stationary and spherically symmetric 
metric of Kerr-Schild form $g = \eta + h \bm{K} \otimes \bm{K}$ such that 
$h$ and $\bm{K}$ are smooth and with no flat zeros can be written in the form
\begin{equation}
 g  = -dT^2 + d\vec{x} \cdot d \vec{x} + h(r) (dr - \sigma dT) \otimes (dr - 
\sigma dT) 
\end{equation}
where $\sigma=\pm 1$ and $r = \sqrt{\vec{x}\cdot \vec{x}}$.
Assume that $h(\vec{x}\,)$ is not of the form
$ h(\vec{x}\,) = \alpha |\vec{x}|^2
\left ( \beta + \gamma |\vec{x}|)^2 \right )^{-1}$
with $\alpha,\beta, \gamma \in \mathbb{R}$, in any domain. Then,
the $g$-geodesic trajectories $(T(s),\vec{x}(s))$ with normalized
tangent vector 
correspond exactly to the solutions of
\begin{align}
\ddot{\vec{x}} &= - \frac{\partial V(r)}{\partial \vec{x}}= 
\frac{\partial}{\partial \vec{x}} \left[ \frac{h(|\vec{x}|)}{2} \left(\frac{L^2}{|\vec{x}|^2} + 
 \mu \right) \right] \label{spatialgeodesic}\\
\vec{L}  & = \vec{x} \times \dot{\vec{x}}  
\label{angularmomentum} 
\end{align}
where $\vec{L}$ is an arbitrary
constant vector, $\mu$  takes the values  $\mu = +1$ for timelike
geodesics, $\mu=0$ for null
geodesics and $\mu=-1$ for spacelike geodesics and
\begin{equation}\label{conservedgeodesic}
\frac{\dot{r}^2}{2}  + \frac{(1 - h(r)) L^2}{2 r^2} - \frac{h(r) \mu}{2} = 
\frac{1}{2} \left( E^2-\mu \right)  := \epsilon
\end{equation}
where $E$ is a constant. Moreover, the tangent vector
$u(s):= (\dot{T}(s),\dot{\vec{x}}(s))$ satisfies
\begin{equation}
E = (1 - h) \dot{T} + h \sigma \frac{\vec{x} \cdot \dot{\vec{x}}}{|\vec{x}|}.
\label{Ener}
\end{equation}

In addition, if the time orientation of $(\M,g)$ is chosen so that
the null vector 
$\partial_T + \sigma \frac{\vec{x}}{|\vec{x}|} \partial_{\vec{x}}$
is future directed, then a geodesic with $\mu=0,1$ 
starting at a point $(T_0,\vec{x}_0 \neq 0)$ is future causal 
if and only if $\dot{\vec{x}}_0$ satisfies (with 
$r_0 := |\vec{x}_0|$, $\dot{r}_0 := \frac{\vec{x}_0 \cdot
\dot{\vec{x}}_0}{|\vec{x}_0|}$ and $h_0 := h(|\vec{x}_0|)$)
 \begin{align*}
& \mbox{if } h_0>1, \quad
\begin{cases}
\begin{aligned}
&\sigma \dot{r}_0 \in [a_0, \infty) \\
&E = \pm \sqrt{\dot{r}_0^2 - a_0^2 }
\end{aligned}
\end{cases}
\hspace{2cm}
\mbox{if } h_0<1,  \quad
\begin{cases}
 \begin{aligned}
& \sigma E \in [a_0, \infty) \\
&\dot{r}_0 = \pm \sqrt{E^2 - a_0^2} \\
\end{aligned}
\end{cases}
\\
& \mbox{if } h_0=1,  \quad  
\begin{cases}  
\begin{aligned}
& \sigma \dot{r}_0 \in [0, \infty)  \quad \quad \mbox{with} \quad
\dot{r}_0 =0 \Longrightarrow
\mu = L =0 \\
&E = \sigma \dot{r}_0
\end{aligned}
\end{cases}
\end{align*}
where $a_0 (r_0,L,\mu) := \sqrt{\left | 1- h(r_0) \right | \left (
\frac{L^2}{r_0^2} + \mu \right )} \geq 0$.
\end{proposition}

\begin{Proof}
The first part of the Proposition is a consequence
of Lemma \ref{Hamiltonianequivalence} in combination
with Remark \ref{completingeqsremark}. Note, in particular,
that (\ref{additional})  (at points where $h(\vec{x})=1$)
must be used to reconstruct
the spacetime trajectory $(T(s),\vec{x}(s))$
from the solutions of 
equations (\ref{spatialgeodesic})-(\ref{angularmomentum}).

For the statements on the initial data, 
let $( T_0, \vec{x}_0 \neq 0)$ be the initial point of the geodesic
and $u_0 = (\dot{T}_0, \dot{\vec{x}}_0)$ the initial velocity, 
normalized to satisfy $g(u_0,u_0) = -\mu$ ($\mu = 0,1$) and assumed
to be future directed.  The initial data 
$(\dot{T}_0, \dot{\vec{x}}_0)$ is equivalent to
$(\dot{T}_0, \vec{L}, \dot{r}_0)$. 
Recall that the Kerr-Schild vector is $K^{\alpha} = (\sigma, 
\frac{\vec{x}}{r})$. The choice of time orientation means
that $\sigma K^{\alpha}$ is future directed. Thus,
$u_0$ being future directed is equivalent to 
$g(u_0, \sigma K |_{s=0}) <0$ or $u_0 = b \sigma K|_{s=0}$, with $b \geq 0$. 
To compute $g(u_0, \sigma K|_{s=0})$ observe that $g(\sigma K, \cdot ) =
 -  dT + \frac{\sigma}{r} \vec{x} \cdot d \vec{x}$ which implies
\begin{equation}
\label{product}
g(u_0, \sigma K|_{s=0} ) = \sigma \dot{r}_0 - \dot{T}_0 .
\end{equation}
On the other hand, the condition $u= b \sigma K|_{s=0}$ ($b \geq 0$)
is ($\dot{T}_0 = b, \dot{\vec{x}}_0 = \frac{b \sigma}{|\vec{x}_0|} \vec{x}_0$)
or equivalently $(\dot{T}_0 = \sigma \dot{r}_0  \geq 0, \vec{L} =0)$. 
Equations (\ref{conservedgeodesic}) and (\ref{Ener}) evaluated at $s=0$ read
\begin{align}
& E^2 = \dot{r}^2_0 + \mbox{sign}(1-h_0) a_0^2, \label{energy2} \\
& E  = \left ( 1-h_0 \right ) \dot{T}_0 + h_0 \sigma \dot{r_0}, \label{E2}
\end{align}
where $\mbox{sign} (1-h_0)$ takes the values $1,0,-1$ depending
on whether $h_0 < 1$, $h_0=1$ or $h_0>1$ respectively
and $a_0$ is as defined in the statement of the Theorem.
At points $h_0 \neq 1$, equations (\ref{energy2})-(\ref{E2})
imply $g(u_0,u_0) = -\mu$.
However, when $h_0=1$, 
(\ref{energy2}) is a trivial consequence
of (\ref{E2}) and $g(u_0,u_0)= - \mu$ must be imposed 
additionally. We compute (with $h_0=1$)
\begin{align}
-\mu & = g (u_0,u_0) = \eta(u_0,u_0) + h_0 \left ( \bm{K} |_{s=0} (u_0) \right 
)^2
= - \dot{T}_0^2 + \dot{\vec{x}}_0^2 + g(\sigma K|_{s=0},u_0)^2 = \nonumber \\
& = 2 \sigma \dot{r}_0 \left ( \sigma \dot{r}_0 - \dot{T}_0 \right ) 
+ \frac{L^2}{r_0^2} ,
\label{caseh0=1}
\end{align}
where (\ref{product}) has been used in the last equality.

We can now find the most general $u_0$ satisfying all these
restrictions. The analysis depends on whether $h_0>1$ $h_0 <1$ 
or $h_0=1$. We start with $h_0 \neq  1$.
Because of (\ref{E2}),
the initial data $\dot{T}_0$  can be substituted by the value of $E$.
Moreover,
\begin{gather*}
(1 - h_0)^2 g(u_0, \sigma K|_{s=0}) = (1-h_0 ) \left ( - ( 1 - h_0 ) \dot{T}_0
+ (1- h_0) \sigma \dot{r}_0 \right ) = \left ( h_0 -1 \right )
\left ( E - \sigma \dot{r}_0 \right ),
\end{gather*}
where (\ref{E2})
has been again inserted in the last equality. 
Thus, the statement $g(u_0, \sigma K_0|_{s=0}) < 0$  or 
$u_0 = b \sigma K|_{s=0}$ with $b \geq 0$ is equivalent to
\begin{gather*}
(h_0 -1 ) (E - \sigma \dot{r}_0 ) < 0 \quad \mbox{ or } \quad 
 E = \sigma \dot{r}_0 \geq  0,
\end{gather*}
the second inequality being a consequence of $\dot{T}_0 = \sigma \dot{r}_0 \geq 0$ and (\ref{E2}). Assume now $h_0 >1$. The conditions to be imposed are 
$\{ E < \sigma \dot{r}_0$ or $E = \sigma \dot{r}_0  \geq 0\}$,
together with $E^2 = \dot{r}_0^2 - a_0^2$ 
(from equation (\ref{energy2})). The locus of this quadratic equation is a hyperbola 
with two branches (degenerating to two straight lines when $a_0=0$)
and with asymptotes $E = \pm \dot{r}_0$. The condition
$\{ E < \sigma \dot{r}_0$ or $E = \sigma \dot{r}_0 \geq 0\} $ selects
precisely the branch satisfying 
$\sigma \dot{r}_0 \geq  a_0$, as claimed in the Proposition.
The case $h_0 <1$ is analogous:  the conditions are now
$\{ E > \sigma \dot{r}_0$ or $E = \sigma \dot{r}_0 \geq 0\}$ together
with
$E^2 = \dot{r}_0^2 + a_0^2$. The solution to these inequalities
is the branch of the hyperbola satisfying 
$\sigma E \geq a_0$.

For the case $h_0=1$, rewrite equation (\ref{caseh0=1})
as
\begin{gather}
2 \sigma \dot{r}_0 \left ( \sigma \dot{r}_0 - \dot{T}_0
\right ) = - \left ( \mu + \frac{L^2}{r_0^2} \right ) \leq 0.
\label{caseh1}
\end{gather}
Thus, the condition
$\{ \sigma \dot{r}_0 - \dot{T}_0 < 0$ or $\dot{T}_0 = \sigma \dot{r}_0 \geq  0\}$
is equivalent to $\sigma \dot{r}_0 \geq 0$ and zero only if
$\mu = L = 0$. This is because, when $\sigma \dot{r}_0 >0$,  
equation (\ref{caseh1}) can be solved uniquely for $\dot{T}_0$ 
with the solution satisfying $\sigma \dot{r}_0 - \dot{T}_0 \leq 0$,
that is, either $\sigma \dot{r}_0 - \dot{T}_0 < 0$ or 
$\dot{T_0} = \sigma \dot{r}_0 > 0$.
When $\sigma \dot{r}_0 =0$ then $\mu=L=0$ and
$\dot{T}_0 \geq 0$ is arbitrary, so again
we satisfy  $\{ \sigma \dot{r}_0 - \dot{T}_0 <0$ 
or $\dot{T}_0 = \sigma \dot{r}_0 \geq 0\} $. 
Finally, the statement $E= \sigma \dot{r}_0$ when $h_0=1$ 
follows directly from (\ref{E2}).
\end{Proof}

\begin{Remark}
Note that when $L=\mu=0$ we have $a_0 \equiv 0$
and this Proposition admits the initial data 
$\dot{r}_0 =0, E=0$ irrespectively
of the value of $h_0$. When $h_0\neq 1$, this boundary case corresponds to the
situation when the initial tangent four-vector 
vanishes, and hence the geodesic is a trivial curve. This is consistent
with the fact that the zero vector is null and future directed.
Admitting trivial curves as null future directed geodesics
has the advantage that allows one to treat at once the cases $\mu =0$ and
$\mu =1 $. 
\end{Remark}

\begin{corollary}\label{epsilonrange}
The variation ranges for $\epsilon$ are
\begin{equation}
\begin{cases}
\epsilon \in [-\frac{\mu}{2},\infty) &\mbox{if } h_0 \geq 1\\
\epsilon \in [\frac{a_0^2-\mu}{2},\infty) &\mbox{if } h_0<1
\end{cases}
\end{equation}
independently of the sign of $\sigma$  and of the function $h(\vec{x})$ in the 
Kerr-Schild metric.
\end{corollary}
\begin{Proof} Immediate from the ranges of variation of $E$
in Proposition \ref{geodesicequationslema} 
and the relation $\epsilon=\frac{1}{2} (E^2-\mu)$.
\end{Proof}

\section{Blow-up of the singularity for radial potentials}\label{McGeheeregularization}

In his original paper \cite{mcgehee1981double}, McGehee proposes a method of
blowing-up the singularity by introducing a coordinate system that 
regularizes the origin for power-law
radial potentials 
$V(|\vec{x}|)=|\vec{x}|^{-\sigma}$ 
in $\mathbb{R}^3$, $\sigma >0$. The field equations are
\begin{equation}
\ddot{\vec{x}}=-\mbox{grad} \left( |\vec{x}|^{-\sigma} \right)= \sigma 
|\vec{x}|^{-\sigma-2} \vec{x}
\end{equation}
where dot is derivative with respect to 
$\tau$ and $\mbox{grad}=\frac{\partial}{\partial x_i}$ is the gradient operator.
Since the trajectories lie in a plane, this system can be restricted to
$\mathbb{R}^2$ without loss of generality. Introducing, as usual,
an auxiliary vector variable $\vec{y}$ this system can be 
rewritten as a first order system on $\mathbb{R}^4$ as 
\begin{align*}
\dot{\vec{x}}&=\vec{y}, \nonumber \\
\dot{\vec{y}}&= \sigma |\vec{x}|^{-\sigma-2} \vec{x}.
\nonumber
\end{align*}
At this point McGehee proposes identifying $\mathbb{R}^2$ with the
complex plane $\mathbb{C}$. Writing $\{ x,y \}$ 
for $\{ \vec{x},\vec{y}\}$ after this identification,
the change of coordinates
\begin{align}
x&= r^{\chi} e^{i \theta} \nonumber \\
y&= r^{-\frac{\sigma}{2} \chi} (u+i v)e^{i \theta},
\label{McGehee}
\end{align}
where $\chi = \frac{2}{2 + \sigma}$
has the two properties of (i) regularizing the system at $r=0$
and (ii)  decoupling the system in the pair of variables $\{u,v\}$.
Thus the original system transforms into an autonomous dynamical system
on the plane $\{ u, v\}$ (with no singularities) together with a pair
of first order ODEs 
in $\{r,\theta\}$ (also free of singularities) which can
be integrated afterwards. The new variables take values $u,v \in \mathbb{R}$,
$r \in [0,\infty)$ and $\theta \in [0,2 \pi)$. 

It is natural to ask whether such a regularization 
and decoupling procedure also occurs  for more general
potentials $V(|\vec{x}|)$. We prove in appendix \ref{nocoordinates}
that only the power-law and the logarithmic potentials
$V(|\vec{x}|) \propto \ln |\vec{x}|$ decouple in the variables 
$\{u,v\}$, even after introducing arbitrary functions of $r$  in the
transformation (\ref{McGehee}). 
Despite this impossibility, the system can still be 
simplified substantially 
by a suitable choice of generalized McGehee transformation.

\begin{theorem}\label{Regularizationtheorem}
Let $\N$ be an open annulus in $\mathbb{C}$
  and $V : \N \rightarrow \mathbb{R}$ be a radially symmetric 
function $V(x) = V(|x|)$. Assume that  $V(|x|)$ is $C^1$ as 
a function of $|x|$ and define $\nabla = \partial_{x^1} + i \partial_{x^2}$
where $x = x^1+ i x^2$, $x^1,x^2 \in \mathbb{R}$. Then the
dynamical system
\begin{align}
\dot{x} &  = y, \nonumber \\
\dot{y} &  = -\nabla V(|x|):= \Lambda(|x|) x, \label{RadialDS}
\end{align}
on $\N \times \mathbb{C}$ 
is equivalent to the
system
\begin{align}
r'& =r  u \nonumber \\
\theta' & =v  \nonumber \\
v'& =-(\beta +1) u \, v \label{TransformedDS}\\
u'& =r^{2-2 \beta } \Lambda (r)-\beta   \nonumber
u^2+v^2,
\end{align}
where $\beta$ is
an arbitrary constant. The coordinates $\{r,\theta,u,v\}$
take values in $r \in (a,b) \subset \mathbb{R^{+}}$,
$\theta \in \mathbb{S}^1$ and $u,v \in \mathbb{R}$. The coordinate change
is defined
by
 \begin{align}
x&= r e^{i \theta} \nonumber \\
y&= r^{\beta} (u+ i v)e^{i \theta} \label{cambiomcgehee} \\
d\tau &= r^{1-\beta} ds, \nonumber
\end{align}
where $\tau$ is the flow parameter in (\ref{RadialDS}) and
$s$ is the flow parameter in (\ref{TransformedDS}).

\end{theorem}

\begin{Remark}
In the case when the potential
$V(|x|)$ is a power-law $V(|x|)= |x|^{-\sigma}$ the transformation
(\ref{cambiomcgehee})
does not agree with the original McGehee transformation (\ref{McGehee})
even after making the choice $\beta = - \frac{\sigma}{2}$. The reason
lies in the specific choice $\chi = 2 /(2 + \sigma)$ made by McGehee.
In fact, any choice of non-zero constant $\chi$ in the transformation
(\ref{McGehee}) 
preserves the same properties for the transformed system. We prefer
to make the choice $\chi =1$ because then $r$ measures directly
the distance of the particle to the origin.
In order to recover the specific form used by McGehee, it is necessary
to apply an additional coordinate change $r \rightarrow r^{\chi}$ 
to the system (\ref{TransformedDS}).
However, this has no benefits for the dynamical system and has the 
drawback of obscuring the clear geometric interpretation
of $r$.
\end{Remark}

\vspace{5mm}

\begin{Proof}
The coordinate change (\ref{cambiomcgehee}) is a
particular case of the coordinate change 
(\ref{GenMcgehee}) introduced in appendix \ref{nocoordinates}
with $\xi_1(r) = r$, $\xi_2(r)=r^\beta$ and $\xi_3 (r)=r^{1-\beta}$.
In particular, equations (\ref{rel1}) and (\ref{rel2}) hold with 
$\alpha = 1$ and $c=1$. Thus, the dynamical system in the 
new coordinates takes the form (\ref{BestDecoupling})
which is exactly (\ref{TransformedDS}) after setting $\alpha=1$, $c=1$.
$\N$ being
an open annulus, it must be of the form
$\N = \{ x \in \mathbb{C} ; x = r e^{i \theta} \mbox{ with }
r  \in (a,b), \theta \in \mathbb{S}^1\}$
for some $0 < a < b$. This proves the claim on the domain of the 
new coordinates. 
\end{Proof}

Concerning the properties of the new dynamical system we have
\begin{theorem}
\label{regularizationtheorem2}
With the same conditions as in Theorem \ref{Regularizationtheorem}, the transformed
dynamical system admits the following two constants of motion
\begin{align}
L & = v r^{\beta +1} \label{constantL} \\
\epsilon & = \frac{1}{2} r^{2\beta} \left (u^2 + v^2 \right ) + V(r).
\label{constantep}
\end{align}
Moreover, if $0 \in \overline{\N}$ and we assume that there
is $\gamma >0$ such that $|x|^{\gamma} V(|x|)$ admits
a $C^1$  extension (as a function of $|x|$) 
to $|x|=0$, then for any $\beta \leq - \frac{\gamma}{2}$ the 
system (\ref{TransformedDS}) admits a $C^0$ extension
to $[0,b)\times \mathbb{S}^1
\times \mathbb{R} \times \mathbb{R}$.
\end{theorem}

\begin{Proof}
The dynamical system (\ref{RadialDS}) describes the motion of a
point particle under the influence of a radial potential $V$.
Thus, the angular momentum $L = \vec{x} \times \dot{\vec{x}}$ 
and the energy $\epsilon = \frac{1}{2} \dot{\vec{x}}^{\,2} + V(\vec{x})$
are conserved. In terms of the complex variables $\{x,y\}$ they take
the form
\cite{mcgehee1981double}:
\begin{align*}
L &= \Im(\bar{x} y)\\
\epsilon &=\frac{1}{2} |y|^2+V(|x|),
\end{align*}
where $\Im$ is the imaginary part and $\bar{x}$ is the complex 
conjugate of $x$. Applying the coordinate change (\ref{cambiomcgehee}) 
one finds
\begin{align*}
L &=v  r ^{1 + \beta} 
\\
\epsilon &=\frac{1}{2} r^{2 \beta } \left(u ^2+v^2\right)+V(r),
\end{align*}
as claimed. Assume now that $\exists \gamma > 0$ such that 
$f(|x|) := |x|^\gamma V(|x|)$ can be $C^1$ extended to $|x|=0$. The function
$\Lambda(|x|)$ (see expression (\ref{RadialDS})) is defined to be
\begin{equation}
\Lambda(|x|)= - \frac{1}{|x|} \frac{dV(|x|)}{d|x|} =
 \gamma \frac{f(|x|)}{|x|^{2+ \gamma}} - \frac{1}{|x|^{1+\gamma}} \frac{df(|x|)}{d|x|}.
\end{equation}
Inserting this into (\ref{TransformedDS}) we see that a sufficient condition
for the dynamical system to admit a $C^0$ extension to $r=0$ is that
$2 \beta + \gamma \leq 0$. 
\end{Proof}

The existence of the first integral $L$ can be used to 
remove $v$ from the equations and reduce the
dimensionality of the system, as well as to
decouple a two-dimensional subsystem.


\begin{lemma}\label{reductionlemma}
The subset of trajectories of Theorem \ref{Regularizationtheorem} with constant
of motion $L$ are equivalent to the dynamical system
\begin{align}
  r'&=r u \label{Generaldecoupled1}\\
u'&=r^{-2 (\beta +1)} \left(L^2+r^4 \Lambda (r)\right)-\beta  
u^2 \label{Generaldecoupled3} \\
 \theta'&=L r^{-(\beta +1)}\label{Generaldecoupled2}
\end{align}
defined on $(a,b) \times S^1 \times \mathbb{R}$.
This system is decoupled in the $\{r,u\}$ variables
and admits the first integral
\begin{equation}
\epsilon=\frac{L^2}{2 r^2}+\frac{1}{2} u^2 r^{2 \beta }+V(r).
\end{equation}
\end{lemma}
\begin{Proof} Solve for $v$ in the constant of motion (\ref{constantL}) and substitute in 
the dynamical system (\ref{TransformedDS}) and in the expression for $\epsilon$
(\ref{constantep}).
\end{Proof}

\subsection{Interpretation of the coordinates}\label{interpretation}

The physical meaning of the coordinates
$\{r,\theta,u,v\}$ follows easily from their definition
(\ref{cambiomcgehee}):
\begin{enumerate}
\item The coordinates $ r,\theta $ are the standard
polar coordinates on the plane.
\item The coordinates $u, v $ are proportional to the radial and the angular 
components of the velocity. This follows from the first equation in (\ref{RadialDS})  because
\begin{align}
r^{\beta} ( u + i v ) e^{i \theta} = \dot{x} = \dot{r} e^{i \theta} + i r e^{i \theta}
\dot{\theta}
\quad \Longleftrightarrow \quad
\left . \begin{array}{ll}
u = r^{-\beta} \dot{r} \\
v =  r^{1- \beta} \dot{\theta}
\end{array}
\right \}.
\end{align}
Thus, $u$ carries all the radial information of the velocity whereas  $v$ encodes the angular part
of the velocity,
\end{enumerate}

The decoupling of the system in the $\{r,u\}$ variables is adequate since it corresponds
to the usual decoupling of the radial motion of a point particle under the
influence of a radial potential. Once this motion is solved, the angular motion
$\theta(s)$ follows by simple integration of $\theta'(s)=L r(s)^{-(\beta +1)}$. 


\subsection{The regularized reduced dynamical system}\label{betaelection}

We have already discussed in Theorem \ref{regularizationtheorem2} the range of values for $\beta$
which regularize the dynamical system 
(\ref{TransformedDS}) at $r=0$. The reduced
dynamical system (\ref{Generaldecoupled1})-(\ref{Generaldecoupled3})
incorporates extra powers of $r$ which are potentially divergent at $r=0$.
The following lemma determines the range of $\beta$ which regularizes
the reduced system.

\begin{corollary}\label{betavalue}
Under the same assumptions as in Theorem \ref{regularizationtheorem2},
the reduced system 
(\ref{Generaldecoupled1})-(\ref{Generaldecoupled3}) for $L\neq 0$
admits a $C^0$ extension to $[0,b) \times \mathbb{S}^1\times \mathbb{R}$
for $\beta \leq \mbox{min} \{ -1 , - \frac{\gamma}{2} \}$.
\end{corollary}
\begin{Proof}
We already know that $\beta \leq - \frac{\gamma}{2}$ regularizes the
term in $\Lambda$ of equation  (\ref{Generaldecoupled3}) at $r=0$. 
For $L\neq 0$, equation
(\ref{Generaldecoupled2}) admits a $C^0$ extension at $r=0$ if
and only if $1 + \beta \leq 0$. This condition also regularizes
the first term in equation (\ref{Generaldecoupled3}). Thus,
$\beta \leq \min \{ -1 , - \frac{\gamma}{2} \}$ is a sufficient 
condition for the existence of a continuous extension to $r=0$.
\end{Proof}


\begin{Remark}\label{remarkoverkill}
The optimal choice of $\beta$
for a detailed study of the dynamical
system (\ref{Generaldecoupled1})-(\ref{Generaldecoupled3}) at $r=0$ is 
$\beta = \mbox{min} \{  -1 , - \frac{\gamma}{2} \}$
with $\gamma$ selected in such a way that $|x|^{\gamma} V(|x|)$ admits a
$C^1$ extension to $|x|=0$ and $\lim_{|x| \rightarrow 0} |x|^{\gamma} V(|x|) \neq 0$.
Indeed, a larger value of $\beta$
is not capable of regularizing the system at $r=0$. On the
other hand, a smaller value of $\beta$
overkills the singularity. This has the effect that the invariant submanifold $\{r=0\}$
(which is called {\it the collision manifold}) has $u=0$ as the
unique
fixed point, and this is always non-hyperbolic. Thus, all details of the phase space
structure of the dynamical system at $\{r=0\}$ are lost by this
choice of $\beta$. 
We will see below an example of this behavior when considering the Schwarzschild limit
of the dynamical system describing causal geodesics in the Reissner-Nordstr\"om spacetime.
\end{Remark}

\section{The Schwarzschild dynamical system}\label{SW}
\begin{figure}[htp!] 
\begin{center}
 \centerline{ \includegraphics[scale=0.5]{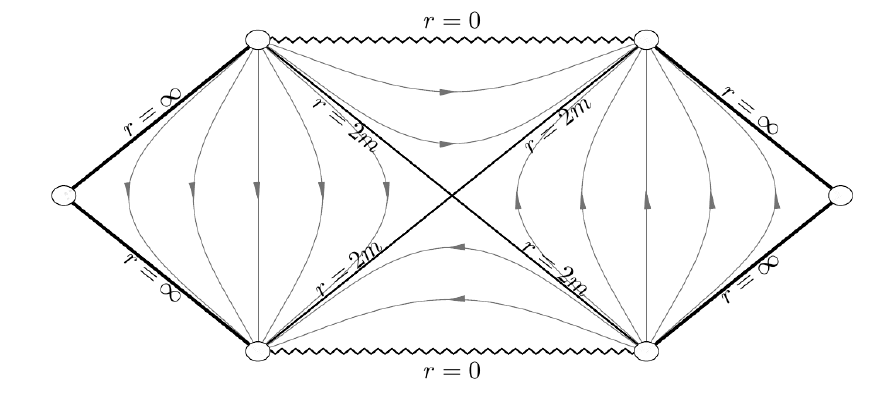}  }
 \end{center}
 \caption{The Schwarzschild Penrose-Carter diagram}
 \label{fig:SWPenrose}
\end{figure}  

As is well-known, the Kruskal spacetime of mass $M >0$ outside 
its bifurcation surface  
can be covered by four patches, two of them isometric
to the advanced Eddington-Finkelstein spacetime and the remaining two
to the retarded Eddignton-Finkelstein. These spacetimes
\cite{1924Natur.113..192E,finkelstein1958past} consist
of the manifold $\mathbb{R} \times \mathbb{R}^+ \times \mathbb{S}^2$
with respective metrics 
\begin{align}
ds^2 & = - \left ( 1- \frac{M}{r} \right ) dV^2 - 2 \sigma dV dr 
+ r^2 d\Omega^2 \label{Edd}
\end{align}
where $\sigma = -1$ for the advanced case and  $\sigma = 1$ for the retarded
one. The coordinates take values in $V \in \mathbb{R},
r \in \mathbb{R}^{+}$ and $d\Omega^2$ is the round unit metric
on the sphere. Furthermore the time orientation is chosen so that
$V$ increases along any timelike curve. The coordinate change
$ V = T - \sigma r $ transforms the metric (\ref{Edd}) into
\begin{equation*}
ds^2 = -dT^2 + dr^2 + r^2 d \Omega^2 + \frac{2M}{r} \left ( dT - \sigma
dr \right )^2
\end{equation*}
with range of variation $ T \in \mathbb{R}$, $r \in \mathbb{R}^+$. Transforming
the flat metric $dr^2 + r^2 d\Omega^2$ to Cartesian coordinates brings the metric
to Kerr-Schild form
\begin{equation}\label{SWmetric}
 g  = -dT^2 + d\vec{x} \cdot d \vec{x} + \frac{2M}{r} (dr - \sigma dT) \otimes 
(dr - \sigma dT) 
\end{equation}
where $r = \sqrt{\vec{x}\cdot \vec{x}}$ and the manifold is
$\mathbb{R} \times (\mathbb{R}^3 \setminus \{ 0 \})$. The choice of
time orientation in (\ref{Edd}) implies that the null vector field
$\partial_T + \sigma \frac{\vec{x}}{|\vec{x}|} \partial_{\vec{x}}$
is future directed.

This form of the metric
was obtained in \cite{kerr1965new}. The case $\sigma = -1$ covers the
upper and right quadrants of the Penrose diagram of the Kruskal
spacetime depicted in Fig. \ref{fig:SWPenrose} and hence approaches the
black hole singularity at $r=0$. The 
case $\sigma = +1$ covers the lower and right quadrants
of the diagram and approaches the white hole singularity at $r=0$. 
Similarly, the spacetime (\ref{SWmetric}) with $\sigma = -1$
also covers the left and upper quadrants of the Kruskal
diagram and the spacetime (\ref{SWmetric}) with $\sigma = 1$
covers the left-lower quadrants.
The only set of points the Kruskal spacetime not
covered by these patches is the bifurcation surface at $r=2M$.

As noticed in Remark \ref{Remarksigma}, the spatial part
of the geodesic equations
do not depend of the choice of sign in $\sigma$ 
and therefore the dynamical system will also be independent
of this choice. This has the following interesting consequence. Consider for
instance a future directed causal geodesic stating
in the region $r < 2M$ in the lower part of the diagram (for simplicity
we call this the {\it white hole region} and by {\it black hole
region} we refer to the domain $r < 2M$ in the
upper part of the diagram). This
geodesic can be described in the Kerr-Schild metric (\ref{SWmetric})
with $\sigma =1$. After a finite value of its affine parameter, the
geodesic will approach $r=2M$. This can 
happen with either $v \rightarrow + \infty$ or $v \rightarrow v_0$ finite.
In either case, since the dynamical system 
only involves the spatial coordinates,
this portion of geodesic will have a limit point $p$ in the phase
space  satisfying $r(p) = 2M$. Irrespectively of which 
spacetime point $q$ is approached
(even it the point lies on the bifurcation surface), the geodesic can 
be continued further as a portion of a geodesic in the spacetime
(\ref{SWmetric}) with $\sigma = -1$ having past endpoint at $q$. 
The fact that the dynamical system for the spatial coordinates
is independent of $\sigma$ implies that the trajectory will be described
in a single phase space, i.e. the change of spacetime chart will pass
fully unnoticed in the phase space of the dynamical system.  Thus, we will
be able to describe the full geodesic as a single trajectory in the 
phase space, without having to determine in which spacetime coordinate
chart  we are working at each portion. As we will see below, this is only
possible due to the presence of a excluded region in the phase space.
In turn, this excluded region arises as a consequence of the restrictions 
in the initial data imposed by the condition that the trajectory 
describes a future directed causal geodesic.

\subsection{Regularized dynamical system}

\begin{lemma}
The McGehee regularization for the dynamical system that describes the spatial 
part of the set of geodesic trajectories with angular momentum
$L$ in the Kruskal spacetime is
\begin{align}
r'&=r u, \label{SistemaSW1}\\
u'&=r \left(L^2-\mu M r\right)-3 L^2 M+\frac{3}{2} u^2,
\label{SistemaSW2}\\
\theta'&=L \sqrt{r}.  \label{SistemaSW3}
\end{align}
where $\mu = 1,0,-1$ for timelike, null and spacelike geodesics, respectively.
The system admits the energy first
integral
\begin{equation}\label{energyfirstintegrealSW}
\epsilon=\frac{u^2}{2 r^3} + \frac{L^2}{2 r^2} -M \left(  
\frac{ L^2 }{ r^3} +
\frac{ \mu}{r}
\right).
\end{equation}
\end{lemma}
\begin{Proof}
We can apply Proposition \ref{geodesicequationslema} with $h = \frac{2M}{r}$,
which corresponds to the spacetime (\ref{SWmetric}). Equations 
(\ref{spatialgeodesic}),  (\ref{angularmomentum}) and (\ref{conservedgeodesic})
become
\begin{align}
\ddot{\vec{x}}&=- \left ( \frac{3 M L^2}{r^5} + \frac{\mu M}{r^3} \right )
\vec{x} , \label{accele} \\
\vec{L}&=\vec{x} \times \dot{\vec{x}}, \nonumber \\
\epsilon:&=
\frac{\dot{r}^2}{2}  +\frac{L^2}{2 r^2} - M \left( \frac{L^2}{r^3} + \frac{ \mu}{ r} \right). \nonumber
\end{align}
where dot means derivative with respect to 
proper time, affine parameter or arc length depending on the value of 
$\mu$ and
 the energy is $E = \dot{T}\left(\frac{2M}{r}-1 \right)+\frac{2M}{r} 
\dot{r}$, from (\ref{Ener}).

Given a geodesic,
we can rotate the Cartesian coordinates so that the geodesic lies
in the plane $\{x^1,x^2\}$ and define $x$
as the complex coordinate $x= x^1 + i x^2$,. The equations of motion 
(\ref{accele}) become
\begin{align*}
\dot{x}&=y,\\
\dot{y}&= - \left ( \frac{3 M L^2}{|x|^5} + \frac{\mu M}{|x|^3} \right ) x 
= - \nabla  \left ( -\frac{ML^2}{|x|^3} - \frac{\mu M}{|x|} \right ) 
:= \Lambda(|x|) x.
\end{align*}
This flow is singular at $r=0$ and we can apply the McGehee regularization
described above. 
From  Corollary \ref{betavalue} and the fact that 
$|x|^3 V(|x|)$ admits a $C^1$ extension to $x=0$ with non-zero value at this
point, we find that
the optimal value for regularization is
$\beta = \mbox{min} \{ -1, - \frac{\gamma}{2} \} = - \frac{3}{2}$.
Applying Lemma \ref{reductionlemma} the following regularized system
is obtained
\begin{align*}
  r'&=r u \\
    u'&=r \left(L^2-\mu M r\right)-3 L^2 M+\frac{3}{2} 
u^2 \\
\theta'&=L \sqrt{r} \\
\epsilon&=\frac{u^2}{2 r^3} + \frac{L^2}{2 r^2} -M \left( 
\frac{L^2}{r^3} + \frac{\mu}{r} \right ).
\end{align*}
\end{Proof}

\subsubsection{Excluded regions}

The dynamical system (\ref{SistemaSW1})-(\ref{SistemaSW3}) describes all future
directed causal
geodesics in the Kruskal spacetime. However, not all trajectories
in this dynamical system correspond to future directed
causal geodesics in this
spacetime. The reason is that the set of initial data $\{ r_0,
\dot{r}_0\}$ for future directed causal geodesics is constrained
by Proposition \ref{geodesicequationslema}.
Given the relation between
$u$ and $\dot{r}$, this implies that not all points $\{r,u,\theta\}$ in the
phase space describe future causal geodesics in the spacetime. 
We will call the {\it allowed region} the set of points in the phase
space corresponding to future directed causal geodesics and the
{\it excluded region} its complement. Let us determine
these sets.

As discussed
in Proposition \ref{geodesicequationslema}, at
points where $h <1$, i.e. $r > 2 M $, there 
is no restriction on the possible values of $\dot{r}_0$, and
consequently no restrictions on $u$ arise.
On the
other hand, when $h > 1$, i.e. $r < 2M$,  then
\begin{equation*}
\sigma \dot{r}_0 \in [a_0, \infty)
\end{equation*}
where
\begin{equation*}
a_0 = \sqrt{\left ( \frac{2M}{r} - 1 \right ) \left ( \frac{L^2}{r^2} 
+ \mu \right )}.
\end{equation*}{
For $h=1$ ($r=2M$), $\dot{r}_0$ is restricted to satisfy
\begin{equation*}
\sigma \dot{r}_0 \in [0 ,\infty)
\end{equation*}
with $\dot{r}_0=0$ only if $L=0$ and the geodesic is null
($\mu=0$). Given that $u = r^{-\beta} \dot{r} = r^{\frac{3}{2}} \dot{r}$ the allowed
region for geodesics in the advanced Eddington-Finkelstein spacetime
(i.e. $\sigma  = -1$) is
\begin{align*}
& u  \in \left (- \infty, - r^{\frac{3}{2}}
\sqrt{\left ( \frac{2M}{r} - 1 \right ) \left ( \frac{L^2}{r^2} 
+ \mu \right )} \, \right ], & r \leq 2M \\
& \mbox{with } (u = 0, r=2M) \mbox{ allowed only if } L=0, \mu =0, \\
\vspace{2mm}
& u  \in (-\infty, \infty ) &  r > 2M 
\end{align*}
So, the excluded region is
\begin{align*}
& u \in \left (-r^{\frac{3}{2}}
\sqrt{\left ( \frac{2M}{r} - 1 \right ) \left ( \frac{L^2}{r^2} 
+ \mu \right )}, \infty \right ), \quad \quad &  r \leq 2M \\
& \mbox{with }( u=0, r = 2M) \mbox{ excluded if } L \neq 0 
\mbox{ or } 
\mu \neq 0  
\end{align*}
Similarly, for geodesics in the retarded Eddington-Finkelstein spacetime
($\sigma =1$) the excluded region is
\begin{align*}
& u  \in \left (- \infty, r^{\frac{3}{2}}
\sqrt{\left ( \frac{2M}{r} - 1 \right ) \left ( \frac{L^2}{r^2} 
+ \mu \right )} \right ), \quad \quad & r \leq 2M \\
&\mbox{with } ( u=0, r = 2M) \mbox{ excluded if } L \neq 0 
\mbox{ or }
\mu \neq 0.
\end{align*}
As discussed above, the bifurcation surface at $r=2M$ is not included
in either the advanced or retarded Eddington-Finkelstein spacetime. 
This is the reason why the the point $(u=0, r=2M)$
when either $L \neq 0$ or $\mu \neq 0$ is excluded. Since future causal 
geodesics with $L \neq 0$
in the Kruskal spacetime do cross the bifurcation surface,
we must incorporate this set of points of the phase space into the allowed
region in order to describe all causal geodesics of the Kruskal spacetime.
We conclude that 
the set of phase space points not describing future directed
causal geodesics in the Kruskal spacetime is the intersection 
of both excluded regions after adding $(u=0,r=2M)$ to the allowed regions,
namely
\begin{align*}
u & \in \left (
- r^{\frac{3}{2}}
\sqrt{\left ( \frac{2M}{r} - 1 \right ) \left ( \frac{L^2}{r^2} 
+ \mu \right )}, 
 r^{\frac{3}{2}}
\sqrt{\left ( \frac{2M}{r} - 1 \right ) \left ( \frac{L^2}{r^2} 
+ \mu \right )} \right ), \quad \quad & r \leq 2M. 
\end{align*}
The existence of these three types of excluded regions is crucial
for describing all future directed causal geodesics
in the Kruskal spacetime in a single phase space diagram. Indeed, any
trajectory passing through an allowed point in the region
$r < 2M$ with $u > 0$ belongs to the $r<2M$ region of a retarded 
Eddington-Finkelstein chart of the  Kruskal spacetime, i.e.
to the white hole region of the spacetime. A trajectory
passing through an allowed point in the region $r < 2M$ with $u  > 0$,
belongs to an advanced Eddignton-Finkelstein chart of the Kruskal spacetime,
i.e. to the black hole region. A point in the phase space with $r > 2M$
belongs to both charts.

The boundary of the allowed region is given by the
set of points satisfying
\begin{eqnarray*}
u^2 = r^3 \left (\frac{2M}{r} -1 \right ) \left ( \frac{L^2}{r^2} + \mu 
\right ), \quad \quad r \leq 2M
\end{eqnarray*}
which can be rewritten as the set of points with $\epsilon = - \frac{\mu}{2}$.
In fact, the excluded region can be equivalently defined as the set
of points for which $\epsilon < - \frac{\mu}{2}$, $r \leq 2 M$, cf.
Corollary \ref{epsilonrange}.

\begin{figure}   
\begin{center}
\psfrag{a}{$r$}  
\psfrag{b}{$u$}
\psfrag{c}{$\sqrt{2M}L)$}
\psfrag{d}{$-\sqrt{2M}L)$}
 \centerline{\includegraphics[scale=0.25]{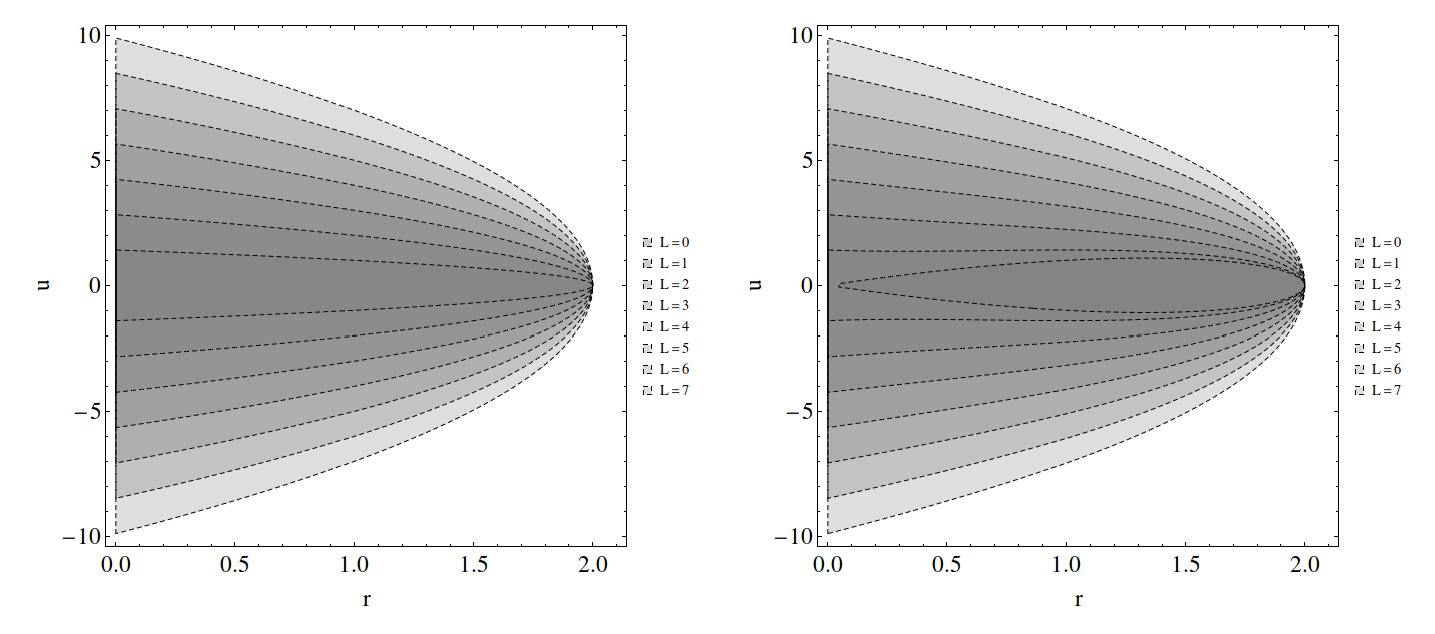}  }
 \end{center}
 \caption{The left image correspond to the excluded regions in the phase space $\{r,u\}$ for null geodesics ($\mu=0$). Different values of the angular momentum $L$ are displayed (the darker the zone, the lower the value of $L$). The right image corresponds to the analogous case for timelike geodesics ($\mu=1$). In both cases
$M=1$.}
 \label{fig:ERegionsSW}
\end{figure}  

The graphical representation of the excluded regions for null and
spacelike geodesics is displayed in Fig. \ref{fig:ERegionsSW}. Notice that,
in the null case, there is no excluded region in the limit $L=0$.
However, in this case the set of points $u=0$
correspond to trivial null geodesics consisting of
single points with vanishing tangent vector.
For non- null geodesics, the excluded region is
always non-empty irrespectively of the value of the angular momentum
$L$.

\FloatBarrier
\subsection{The collision manifold}\label{collisionmanifold}
\begin{figure}   
\begin{center}
 \centerline{\hspace{0.1\textwidth} \includegraphics[scale=0.225]{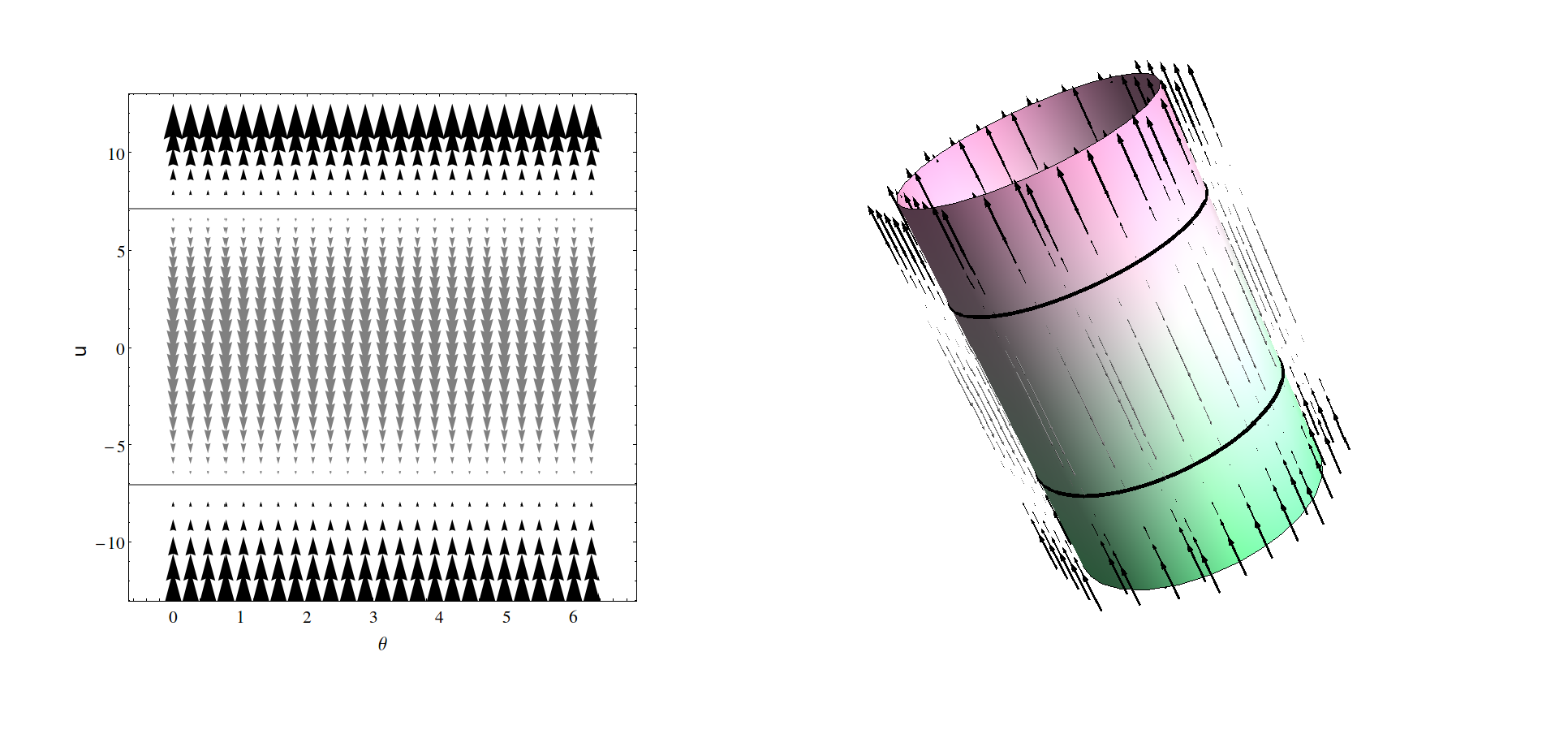}  }
 \end{center}
 \caption{The flow in the collision manifold (left) with the critical lines in 
  $u=\pm \sqrt{2M}L$ (we have chosen $L=5 ,M=1$) and the collision 
manifold embedded cylinder (right) with 
the flow
and the critical lines over it.}
 \label{fig:collisionSW}
\end{figure}  

The submanifold $r=0$ is clearly invariant under the flow. Since $r=0$
corresponds to the spacetime singularity, this submanifold is called
{\it collision manifold}. It can be described
globally by the coordinates $\{(\theta,u)\}$ so its topology is
$\mathbb{R} \times \mathbb{S}^1$.
The dynamical system (\ref{SistemaSW2})-(\ref{SistemaSW3}) 
restricted to the collision manifold reads
  \begin{align*}
  u'&=\frac{3}{2} u^2-3 L^2 M, \\
\theta'&=0.
\end{align*}
This system has two lines of critical 
points: one line of stable points at $(\theta,u)=(\theta_0,-\sqrt{2M L^2})$ 
and one line of unstable nodes at $(\theta,u)=(\theta_0,\sqrt{2ML^2})$, where 
$\theta_0 \in \mathbb{S}^1$ is an arbitrary value.
The phase space
portrait in the collision manifold is shown 
in Fig. \ref{fig:collisionSW}. For each value of $\theta_0$, there is a 
trajectory extending from $u = - \infty$ and approaching $u=-
\sqrt{2 M L^2}$ as its future limit point, a trajectory from 
$u=- \sqrt{2 M L^2}$ to $u= \sqrt{2M L^2}$ and a trajectory
having $u= \sqrt{2 ML^2}$ as its past limit point and extending
to $u = + \infty$, all of them with $\theta= \theta_0$. One may wonder
how these trajectories relate to causal geodesics in the Kruskal
spacetime. To see this, note that any such geodesics must have a finite
value of $\epsilon$. On the other hand $\epsilon$ diverges at $r=0$ (see
(\ref{energyfirstintegrealSW})). In fact, it does so in the following way
\begin{eqnarray*}
\lim_{r \rightarrow 0} \epsilon = \left \{ 
\begin{array}{l}
+ \infty 
\quad \quad \mbox{if} \quad  |u| >  \sqrt{2 M L^2}, \\
- \infty 
\quad \quad \mbox{if} \quad |u| <  \sqrt{2 M L^2}. \\
\end{array}
\right .
\end{eqnarray*}
The limit when $|u|$ approaches the critical points on the collision
manifold depends on the details of how this limit is taken. Since,
the 
trajectories joining the stable critical points to the unstable
ones within the collision manifold 
are interior to the excluded region of the phase phase, it follows
that such trajectories are completely unrelated to 
causal geodesics in the spacetime. This is consistent with the fact that
$\epsilon \rightarrow - \infty$ on these trajectories,
while future directed causal geodesics in the region $r < 2M$ must have
$\epsilon \geq - \frac{\mu}{2}$.
On the other hand, the trajectories on  the collision manifold
leaving $ u = + \sqrt{2M L^2}$ and those approaching
$ u = - \sqrt{2M L^2}$ correspond to the limit of trajectories 
of causal geodesics leaving the white hole  singularity
at $r=0$ and approaching the black hole singularity at $r=0$
when their energy $\epsilon$ diverges to $+ \infty$. Thus, this set
of trajectories on the collision manifold carries interesting information
on the causal geodesics in the spacetime.

To analyze the behavior of the particles near the 
collision manifold, we can linearize the system at its first order as
\begin{align*}
r(s)&=\delta r(s),\\
u(s)&=u_0(s)+\delta u(s),
\end{align*}
where $u_0(s)$ its a solution that corresponds to an orbit in the collision 
manifold and therefore satisfies
\begin{equation*}
u_0'=-3 L^2 M+\frac{3 u_0^2}{2}.
\end{equation*}
The general solution for this differential equation satisfying $|u_0| > 
\sqrt{2ML^2}$ is
\begin{equation*}
u_0(s)=- \sqrt{2 M L^2} \coth \left( 3 \sqrt{\frac{M L^2}{2}} s\right).
\end{equation*}
The branch $s\in (- \infty,0)$ corresponds to the solution leaving the unstable 
fixed point at $u=  + \sqrt{2ML^2}$  and approaching 
$u \longrightarrow + \infty$, while the branch $s \in (0, + \infty)$
corresponds to the solution extending from $u =- \infty$ and
approaching the fixed point $u = - \sqrt{2 ML^2}$.
The first-order linearized system in the variables  $\delta r$ and
$\delta u$ is 
\begin{align*}
\delta r' &= (\delta r) u_0 ,\\
\delta u' &=L^2 (\delta r) + 3 (\delta u) u_0 .
\end{align*}
We can easily solve for $\delta r(s)$:
\begin{equation*}
\delta r(s)=\frac{\delta r_0}{\left |\sinh \left(
3 \sqrt{\frac{ML^2}{2}} s \right ) \right |^{\frac{2}{3}}},
\end{equation*}
where $\delta r_0>0$ is an arbitrary integration constant.
Thus, the fixed points are approached exponentially fast in the variable $s$.
To analyze the behavior in the $\tau$ variable we recall the relation
(\ref{cambiomcgehee}), namely
\begin{equation*}
 \tau'(s)=(\delta r(s))^{\frac{5}{2}}.
\end{equation*}
This integration can be performed explicitly, but it is not particularly
enlightening. Instead, we will determine a series expansion of
$\delta r(\tau)$ near $\tau = 0$  where $\tau$ is chosen
so that the particle reaches
the singularity $r=0$ at $\tau =0$ (note that $\tau <0$
for particles approaching the black hole singularity
while $\tau >0$ for particles
leaving the white hole singularity). Define $x(s)$ as
\begin{equation*}
x(s)= \frac{1}{\left |\sinh \left (3 \sqrt{\frac{ML^2}{2}} \,  s \right ) \right |^{\frac{2}{3}}}
\end{equation*}
so that $(\delta r)(s) = (\delta r_0) x(s)$ and hence
$\frac{dx}{ds} = u_0 x$. In terms of $x$,
\begin{equation*}
u_0 = \mp  \sqrt{2M L^2} \sqrt{1+ x^3},
\end{equation*}
where the sign depends on the branch we are considering (upper
sign for the approach to the black hole and lower sign for the white
hole). Then
\begin{equation*}
\frac{d \tau}{dx} = \frac{d \tau}{ds} \frac{ds}{dx} = \frac{(\delta r_0 
\, x)^{\frac{5}{2}} }{x u_0} = 
\mp \frac{(\delta r_0)^{\frac{5}{2}}}{\sqrt{2ML^2}}
\frac{x^{\frac{3}{2}}}{\sqrt{1+ x^3}}.
\end{equation*}
Expanding the right-hand side as a series en $x$, integrating and inverting
we find 
\begin{equation*}
x(\tau)= \left ( \frac{5}{2} \right )^{\frac{2}{5}}
\left ( \mp a \tau
\right )^{\frac{2}{5}}+ \frac{5}{22}
\left ( \frac{5}{2} \right )^{\frac{3}{2}}
\left ( \mp a \tau 
\right )^{\frac{8}{5}} + O(\tau^{\frac{18}{5}}) + \dots
\end{equation*}
where $a = \frac{\sqrt{2 ML^2}}{(\delta r_0)^{\frac{5}{2}}}$.
A plot of the function $\delta r(\tau)$ in each of the two branches is given in Fig. \ref{fig:CollisionSW}. These plots describe
the approach to the singularity of very energetic particles 
in the Kruskal spacetime for different values of the angular momentum.
Note that 
the result is the same for massive and massless particles,
as one could expect given the very high energy involved. 
  
  \begin{figure}   
\begin{center}
 \centerline{\includegraphics[scale=0.3]{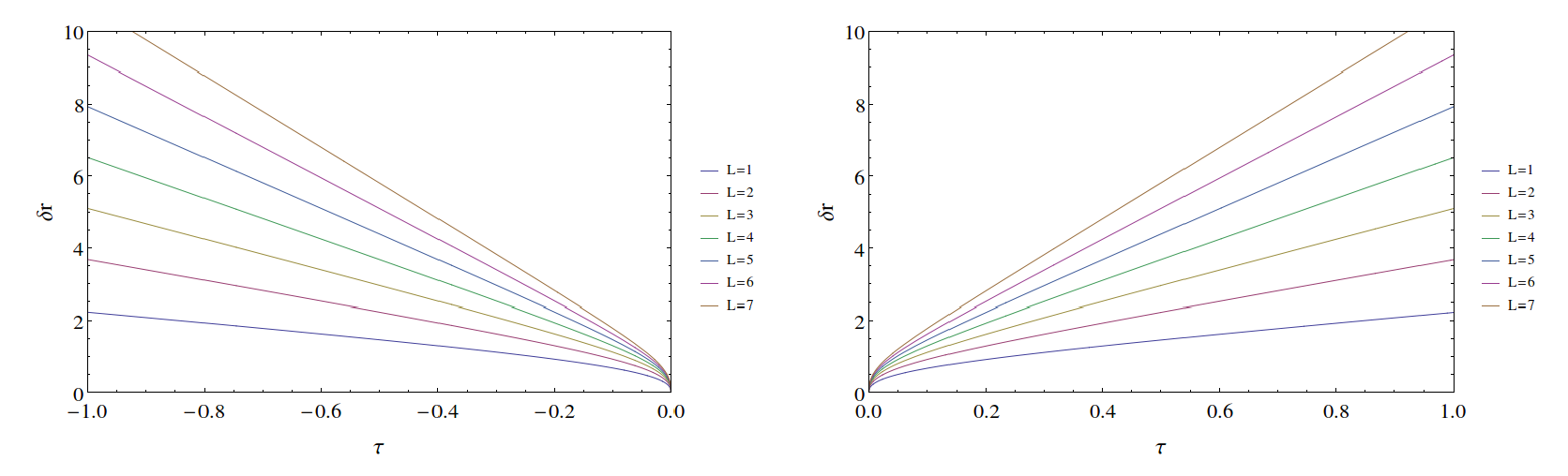}}
 \end{center}
 \caption{Plot of the function $\delta r(\tau)$. The image on the left corresponds to the branch in witch $u_0\leq -\sqrt{2 M L^2}$ (black hole solution) in the collision manifold and the image on the right corresponds to the branch in witch $u_0\geq \sqrt{2 M L^2}$ (white hole solution). In both plots $\delta r_0=1$ and $M=1$.}
 \label{fig:CollisionSW}
\end{figure}  
  \FloatBarrier
\subsection{The general flow}

\subsubsection{Massless particles}

\begin{figure}
\centering
\begin{tabular}{cc}
\includegraphics[scale=0.225]{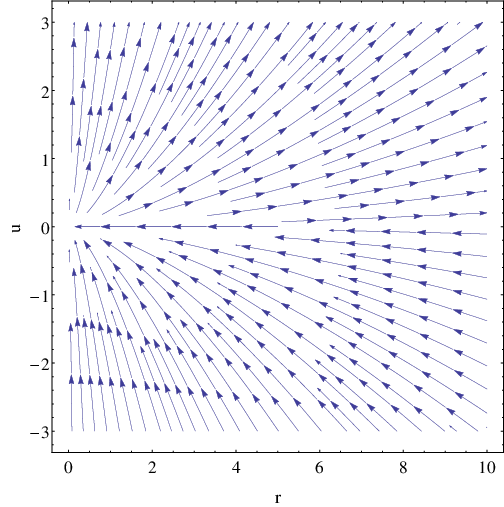}&
\includegraphics[scale=0.225]{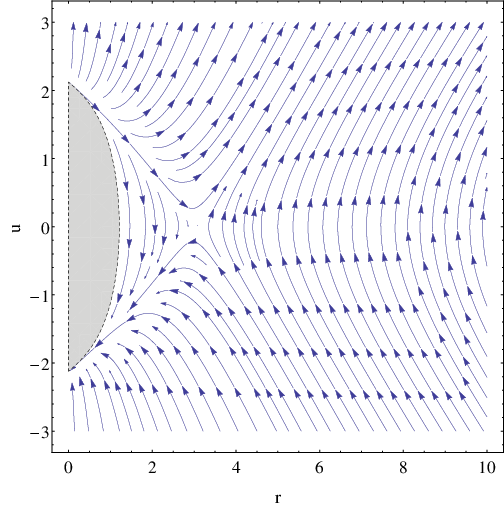}
\end{tabular}
\vspace*{1em}
\caption{Phase space for massless particles with $ L = 0 $ (left picture) and $ L = 1.5 M $ (right picture). The dark zone correspond to the forbidden region given by $\epsilon<\frac{\mu}{2}$.}
  \label{fig:Fullphasespacelight}
\end{figure} 
The reduced dynamical system with $\mu=0$ takes the form
\begin{align}\label{SistemaSWluz}
r'&=r \, u,\\
u'&=r L^2-3 L^2 M+\frac{3 u^2}{2}.
\end{align}
Its phase space is displayed for different values of $L$
in Fig. \ref{fig:Fullphasespacelight}. The fixed points are (assuming
$L \neq 0$)
\begin{equation*}
(r=0, u =\pm \sqrt{2M} L) \quad \quad 
(r =3M, u=0).
\end{equation*}
The linearization of the system around these points has the following
eigenvalues
\begin{equation*}
\begin{cases}
\lambda_1 = \sqrt{3 M L^2} , \quad
\lambda_2 = - \sqrt{3 M L^2} &\mbox{for } \quad  (r=3M, u=0),\\
\lambda_1=\pm 3 \sqrt{2M L^2} \quad \lambda_2=\pm \sqrt{2M L^2}  &\mbox{for} 
\quad (r=0, u = \pm \sqrt{2M} L).
\end{cases}
\end{equation*}
Thus, all critical points are hyperbolic. The
point $(r=0,u=+ \sqrt{2ML^2})$ is an unstable node (repulsor),
$(r=0,u=- \sqrt{2ML^2})$ is a stable node (attractor)
and $(r=3M,u=0)$ is saddle point. This saddle point obviously corresponds
to the unstable circular orbit  for massless particles.

\subsubsection{Massive particles}

\begin{figure}
\centering
\centerline{
\begin{tabular}{cccc}
\includegraphics[scale=0.225]{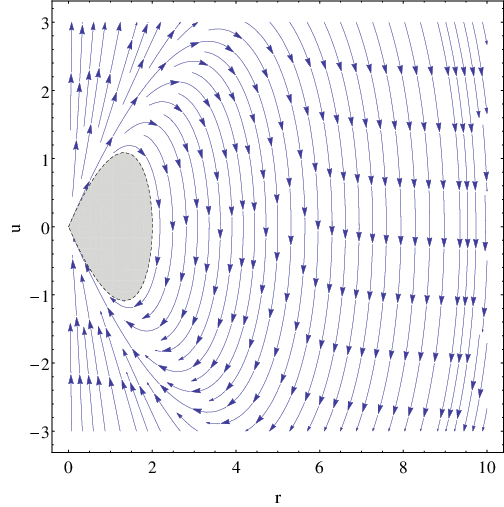}&
\includegraphics[scale=0.225]{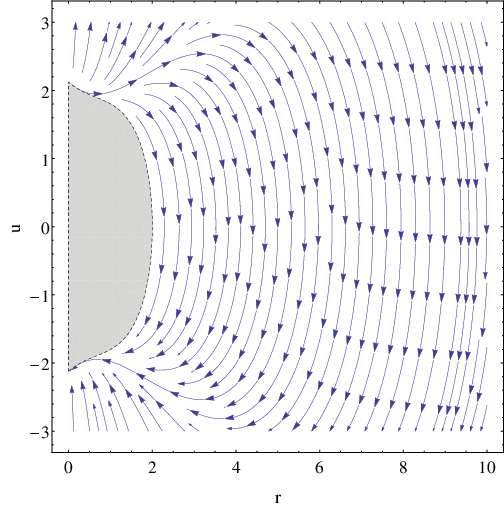}&
\includegraphics[scale=0.225]{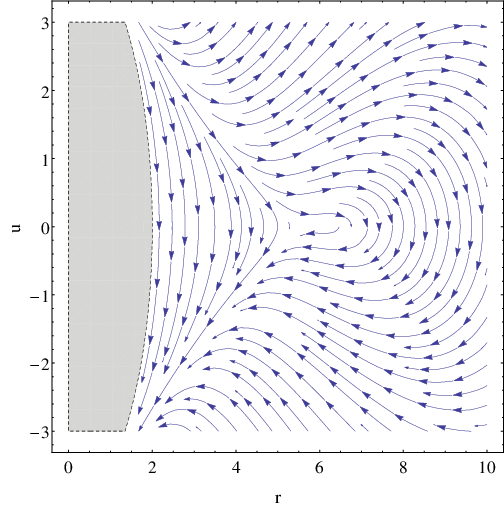} &
\includegraphics[scale=0.225]{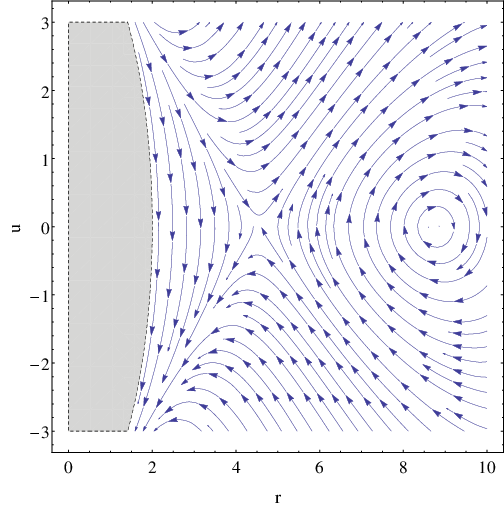}
\end{tabular}}
\vspace*{1em}
\caption{Phase spaces for massive particles with  $ L = 0 $, 
$ L = 1.5 M $, $ L = 2 \sqrt {3} M$ and $L = 3.65 M $.
The dark zone correspond to the forbidden region given by
$\epsilon<\frac{\mu}{2}$. The larger the value of $L/M$ the larger the
excluded region in the phase space.}
  \label{fig:Fullphasespace}
\end{figure}
When $\mu=1$ the reduced dynamical system is
\begin{align*}
r'&=r \, u,\\
u'&=r \left(L^2-M r\right)-3 L^2 M+\frac{3 u^2}{2},
\end{align*}
with phase spaces displayed for different values of $L$ in Fig.
\ref{fig:Fullphasespace}. The fixed points are
\begin{equation*}
(r=0, u =\pm \sqrt{2M L^2}), \quad \quad 
(r = r_{\pm}(M,|L|) := \frac{L^2 \pm  |L|\sqrt{L^2-12 M^2}}{2 M}, u=0),
\end{equation*}
the second pair under the additional condition $L^2 \geq 12 M^2$. 
For $L^2 > 12 M^2$ all fixed points are hyperbolic, with $ (r= r_+(M,|L|), u=0)$
being a center (purely imaginary eigenvalues) and $(r=r_{-}(M,|L|), u=0)$
being a saddle. When $L^2 = 12 M^2$, there is a bifurcation in the phase
space, which can be visualized in the transition between the
third and fourth plots in Fig. \ref{fig:Fullphasespace}. Given that $r_+(M,|L|)$
is an increasing function of $|L|$ with values ranging from $(6M,\infty)$
we recover the well-known fact that the innermost stable circular
orbit (ISCO) in Schwarzschild is $r= 6M$. The fixed point $r_{-}(M,|L|)$
is a decreasing function of $|L|$ with values ranging from $6M$ (when $|L|
\rightarrow 2 \sqrt{3} M$) to $3M$ when $|L| \rightarrow + \infty$).
We thus recover easily all well-known results for geodesics in Schwarzschild
outside the horizon. The approach here, however, is perfectly regular both
across the horizon at $r=2M$ and even at the singularity $r=0$. Moreover,
it allows us to treat all points in the Kruskal spacetime with a single
dynamical system.

\FloatBarrier

\section{The Reissner–Nordstr\"om dynamical system}\label{RN}

The Reissner–Nordstr\"om spacetime
\cite{reissner1916eigengravitation,nordstrom1918energy} corresponds to
the most general solution for Einstein equations with electromagnetic
field when spherical symmetry is assumed. The maximal extension of
this spacetime is covered by a infinite amount of copies of four basic
patches. As in the Kruskal spacetime two of the patches are  isometric
to the Reissner–Nordstr\"om advanced Eddington-Finkelstein spacetime and
the remaining two to the Reissner–Nordstr\"om retarded
Eddignton-Finkelstein. Each one of these spacetimes consist of the
manifold  $\mathbb{R} \times \mathbb{R}^+ \times \mathbb{S}^2$ with
respective metrics
\begin{equation*} ds^2 = - \left ( 1- \frac{2M}{r} + \frac{Q^2}{r^2}
\right ) dV^2 -2 \sigma dV dr + r^2 d \Omega^2,
\end{equation*} 
where $\sigma = -1$ for the advanced case and  $\sigma
= 1$ for the retarded one. The coordinates takes values in $V \in
\mathbb{R},r \in \mathbb{R}^{+}$ and $d\Omega^2$ is the round unit
metric on the sphere. 
Performing the same coordinate change as in the Kruskal case,
the metric is transformed into its Kerr-Schild
form \cite{kerr1965new}:

\begin{equation}\label{RNmetric}
 g  = -dT^2 + d\vec{x} \cdot d \vec{x} + \frac{2M r - Q^2}{r^2} (dr - \sigma dT) \otimes 
(dr - \sigma dT) 
\end{equation}
where $r = \sqrt{\vec{x}\cdot \vec{x}}$ and the manifold is $\mathbb{R} \times (\mathbb{R}^3 \setminus \{ 0 \})$. As before we choose the orientation
so that the nowhere-zero, null vector 
$\partial_T + \sigma \frac{\vec{x}}{|\vec{x}|} \partial_{\vec{x}}$ is
future directed.
\begin{figure}[htp!] 
\begin{center}
 \centerline{ \includegraphics[scale=0.5]{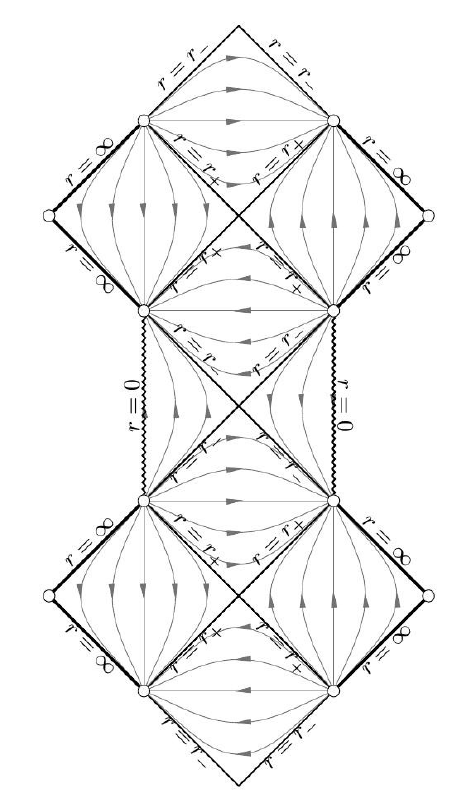}  }
 \end{center}
 \caption{The Reissner–Nordstr\"om Penrose-Carter diagram.}
 \label{fig:RN}
\end{figure}   

As is well-known,
the global structure of the maximal extension of the Reissner-Nordstr\"om
spacetime depends strongly on whether $|Q| > M$, $|Q| = M$ or $|Q| < M$.
For the discussion below we concentrate on the latter case because
it corresponds to a non-extremal black hole. We emphasize, however,
that all the dynamical
systems in this Section are valid for any value of $Q$ and $M$, so they can be used
to study geodesics in the extremal black hole or naked singularity cases
as well.

When $|Q| < M$, the basic structure of the maximal
extension of the Reissner–Nordstr\"om spacetime has now two bifurcation
surfaces located, respectively,
at the intersection of the two smooth null
hypersurfaces with $r=r_+$ and the two smooth null hypersurfaces
with $r=r_{-}$, where $r_{\pm}=M\pm\sqrt{M^2-Q^2}$. We call each one of these
four hypersurfaces a {\bf horizon}. The patch with $\sigma
= -1$ covers the structure unit that goes in ascending way
from $r=\infty \to r=r_+ \to
r=r_- \to r=0$ (starting from either the right or the left)
while the patch with $\sigma = 1$ covers the structure unit that goes
in an ascending way from $r=0 \to r=r_- \to r= r_+ \to  r=\infty$  
(starting from the right or the left). Notice that there is no
causal geodesic starting from a left-most quadrant 
which, after  crossing 
the null hypersurface $r=r_+$ then
goes across the portion of the null hypersurface at $r=r_-$ 
lying at the left of the diagram (and similarly for geodesics starting
at a  right-most quadrant). The
reason is that  the Killing 1-form 
$\bm{\xi} := g(\xi,\cdot)$ where $\xi = \partial_T$ 
(defined on a single Eddington-Finkelstein patch) is integrable with
orthogonal hypersurfaces foliating the region $r_{-} < r < r_{+}$
with timelike leaves. Let us label these leaves by $t$. As a
consequence we have $\bm{\xi} = G d t$ on $r_{-} < r < r_{+}$
where $G$ is a smooth function.
Consider the 
conserved energy for the geodesic, i.e. $ \langle \partial_T, u \rangle =-E $
where $u$ is the tangent vector. In the region between
$r_-$ and $ r_+$, in order for the geodesic to enter from the left portion of
the null hypersurface $r= r_{+}$ and leave across the left portion
of the hypersurface $r=r_{-}$, the geodesic must become somewhere tangent
to a hypersurface $t= \mbox{const}$. At this point we have $-E = \bm{\xi} (u) =
G dt (u) = 0$.
But $E=0$ is impossible for a geodesic starting on the left-most region
where $\xi$ is timelike. A similar argument applies to geodesics
starting at the right-most quadrant.

As discussed in Section \ref{SW}, the fact that the dynamical system for the
spatial coordinates is independent of $\sigma$ implies that the
trajectory will be described in a single phase space. To understand
the geodesic flow along the maximal extension of the
Reissner–Nordstr\"om we need to analyze the excluded regions.
We first write down the regularized dynamical system.

\subsection{Regularized dynamical system}
\label{regularizedRN}
\begin{lemma}
The McGehee regularization for the dynamical system that describes the spatial 
part of the set of geodesic trajectories with angular momentum
$L$ in the Kruskal spacetime is
\begin{align}
r'&=r u, \label{SistemaRN1}\\
\theta'&=L \, r, \label{SistemaRN2}\\
u'&=r \left(r \left(L^2-\mu  M r+\mu  Q^2\right)-3 L^2 M\right)+2 \left(L^2 Q^2+u^2\right). \label{SistemaRN3}
\end{align}
where $\mu = 1,0,-1$ for timelike, null and spacelike geodesics, respectively.
The system admits the energy first
integral
\begin{equation}
\label{energyfirstintegrealRN}
L^2 \left(r (r-2 M)+Q^2\right)+\mu  r^2 \left(Q^2-2 M r\right)+u^2=2 r^4 \epsilon.
\end{equation}
\end{lemma}
\begin{Proof}
We can apply Proposition \ref{geodesicequationslema} with $h = \frac{2M r-Q^2}{r^2}$, from (\ref{RNmetric}). Equations 
(\ref{spatialgeodesic}),  (\ref{angularmomentum}) and (\ref{conservedgeodesic})
become
\begin{align*}
\ddot{\vec{x}}&=\left( \left ( \frac{2Q^2L^2 }{r^6} + \frac{\mu Q^2}{r^4} \right )  -\left ( \frac{3 M L^2}{r^5} + \frac{\mu M}{r^3} \right ) \right)
\vec{x} , \nonumber \\
\vec{L}&=\vec{x} \times \dot{\vec{x}}, \nonumber \\
\epsilon:&=
\frac{\dot{r}^2}{2}  +\frac{L^2}{2 r^2} - M \left( \frac{L^2}{r^3} + \frac{ \mu}{ r} \right) + Q^2 \left( \frac{L^2}{2 r^4} + \frac{\mu }{2 r^2} \right). 
\nonumber
\end{align*}
Adapting coordinates so that the geodesic lies in the $\{ x^1,x^2\}$
plane and defining the complex variable  $x= x^1 + i x^2$, the equations
are 
\begin{align*}
\dot{x}&=y,\\
\dot{y}&=\left( \left ( \frac{2Q^2L^2 }{|x|^6} + \frac{\mu Q^2}{|x|^4} \right )  -\left ( \frac{3 M L^2}{|x|^5} + \frac{\mu M}{|x|^3} \right ) \right) x
\\
&= - \nabla  \left ( -M \left( \frac{L^2}{|x|^3} + \frac{ \mu}{|x|} \right) +
Q^2 \left( \frac{L^2}{2 |x|^4} + \frac{\mu }{2 |x|^2} \right) \right ) :=
\Lambda(|x|) x. 
\end{align*}
We now apply the McGehee regularization. Since
$|x|^4 V(|x|)$ admits a $C^1$ extension to $x=0$,
Corollary \ref{betavalue} fixes the optimal value for regularization as
$\beta = \mbox{min} \{ -1, - \frac{\gamma}{2} \} = - 2$
and Lemma \ref{reductionlemma} gives the regularized system 
(\ref{SistemaRN1})-(\ref{SistemaRN3})
as well as the constant of motion $\epsilon$. \end{Proof}

\subsubsection{Excluded regions}\label{excludedRN}

As in the Schwarzschild case, not all trajectories of the dynamical
system (\ref{SistemaRN1})-(\ref{SistemaRN3}) 
correspond to future directed
causal geodesics in the Reissner-Nordstr\"om spacetime.  Proceeding
as in Schwarz\-schild and exploiting 
 Proposition \ref{geodesicequationslema}, it is straightforward to
find that the excluded region of the phase space diagram is
\begin{align*}
u & \in \left (
- r^{2}
\sqrt{\left (  \frac{2M}{r}-\frac{Q^2}{r^2} - 1 \right ) \left ( \frac{L^2}{r^2} 
+ \mu \right )}, 
 r^{2}
\sqrt{\left (  \frac{2M}{r}-\frac{Q^2}{r^2} - 1 \right ) \left ( \frac{L^2}{r^2} 
+ \mu \right )} \right ), \quad \quad &  r_-\leq r \leq r_+ . 
\end{align*}
In addition, for geodesics in the $\sigma =-1$ Eddington-Finkelstein patch
$u$ must be lie below the forbidden region (i.e. $u \leq 0$ in 
the strip $r_{-} \leq r \leq r_+$),
while geodesics in the $\sigma =+1$ Eddington-Finkelstein patch 
must lie above the forbidden region (i.e. $u \geq 0$ in the strip
$r_{-} \leq r \leq r_{+}$ ). Note that, as in Schwarzschild,
the boundary of the allowed region is defined by
\begin{eqnarray*}
u^2 = r^4 \left (\frac{2M}{r}-\frac{Q^2}{r^2}  -1 \right ) \left ( \frac{L^2}{r^2} + \mu 
\right ), \quad \quad r_-\leq r \leq r_+,
\end{eqnarray*}
which correspond to the set of points with $\epsilon = - \frac{\mu}{2}$
and the excluded region is precisely the set
$\epsilon < - \frac{\mu}{2}$, $r_-\leq r \leq r_+$, cf. Corollary \ref{epsilonrange}. 

 The excluded regions for timelike geodesics are displayed in Fig. \ref{fig:collision_RN} with representative values of $Q$ and $L$. The excluded regions for 
timelike/null geodesics show the same behavior on $L$ as in the Schwarzschild case,
i.e. in the null case there is no excluded region in the limit $L=0$ but then
the line $u=0$ consists of a family of trivial geodesics and
in the timelike case the excluded region is always non-empty for all
values of $L$.

We can now discuss how the behavior of geodesics across the
horizons can be described in the phase space diagram. 
The crucial information is the restriction for
$u$ (which, recall, is proportional to $\dot{r}$)
in the domain $r_-<r < r_+$. Let us see how
this implies that any geodesic traveling
from $r=r_0> r_+ \to r_+ \to r_{-} \to r_1$ and back to $r_- \to r_+$ 
must have changed the Kerr-Schild patch
along the way (by changing the value of
$\sigma$). Assume for definiteness that the particle starts
in a left-most quadrant. After the crossing of $r_+$ the particle
must necessarily cross $r=r_{-}$. This well-known fact can be 
directly deduced also from the dynamical system because the allowed region
in the domain $r_{-} < r < r_{+}$ when $\sigma = -1$ has $u <0$ everywhere.
The crossing of $r= r_{-}$ happens in the right part of the 
Kerr-Schild patch, as discussed before.
Thus the
trajectory is still contained in the original Kerr-Schild
patch and, at the same time, it has also entered a new patch
with $\sigma = -1$.  Since the collision manifold
cannot be attained (see below)
the geodesic must necessarily reach a point where 
$u=0$ and cross to positive values of $u$. From then on, the curve
crosses $r = r_{-}$ towards larger values of $r$. At this crossing, the
trajectory necessarily leaves the original Kerr-Schild patch. 
It does so with $u>0$
which is compatible with the fact that we now have $\sigma=-1$ and
hence the allowed region lies above the forbidden bubble. The 
geodesic  necessarily crosses $r=r_{+}$ and enters
a different asymptotic region from which it started. This behavior
is of course well-known. What is important here is that
a single phase space allows for a complete
description of the complicated spacetime trajectory by simply noticing that
each time that the forbidden region is encircled, we have moved one step
in the ladder in Kerr-Schild patches in the maximal
extension of Reissner-Nordstr\"om. This is possible only because
(i) the phase space has an excluded region and (ii) the crossing
at $r=r_{-}$ is not ambiguous, forcing the particle to stay
in the same Kerr-Schild patch it started.
\begin{figure}
\centering
\begin{center}
\centerline{
\includegraphics[scale=0.425]{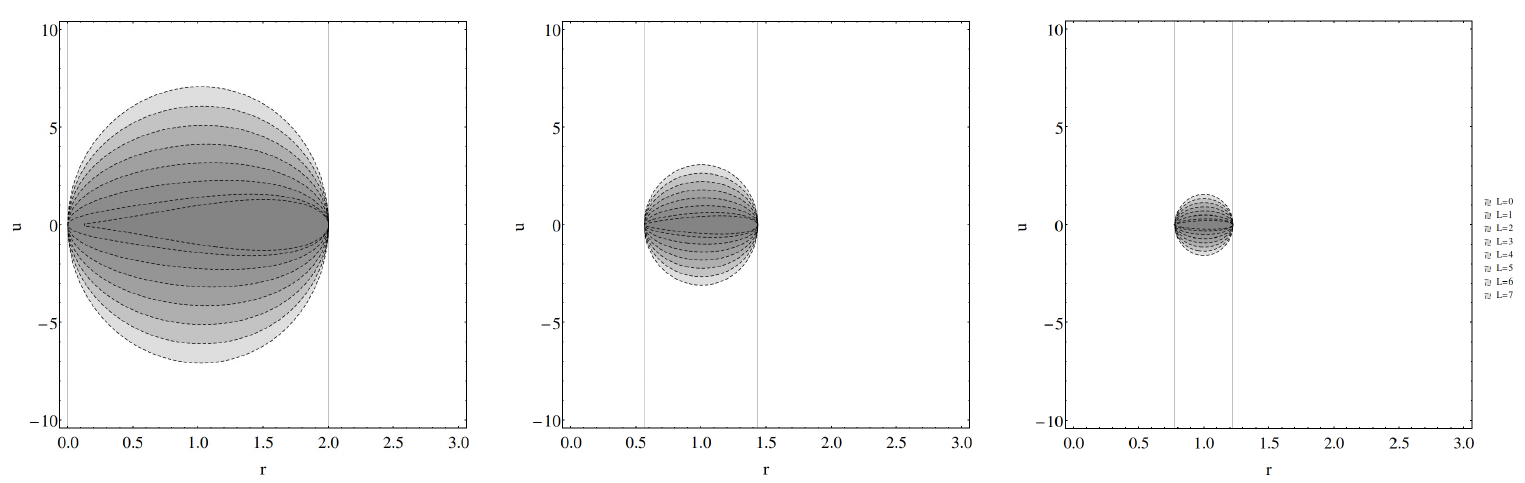}
}
\end{center}
\vspace*{1em}
\caption{The three pictures displayed correspond to the excluded regions in the phase space $\{r,u\}$ for timelike geodesics ($\mu=1$) with $Q=0$, $Q=0.9$ and $Q=0.975$ respectively. Different values of the angular momentum $L$ are displayed in each one (the darker the zone, the lower the value of $L$). Units
are chosen so that $M=1$. The vertical lines correspond to $r_\pm$.}
  \label{fig:collision_RN}
\end{figure}
\subsection{The collision manifold}\label{RNcollisionmanifold}
The Reissner-Nordstr\"om collision manifold has
again  topology $\mathbb{R} \times \mathbb{S}^1$ and global
coordinates $\{(\theta,u)\}$.
The dynamical system (\ref{SistemaRN2})-(\ref{SistemaRN3}) 
restricted to the collision manifold reads
  \begin{align}
  u'&=2 \left( L^2 Q^2+u^2 \right), \nonumber \\
\theta'&=0, \label{colisionRN}
\end{align}
which has no fixed points. This
is a manifestation of the fact that, in Reissner-Nordstr\"om,
the singularity has a
repulsive behavior,  unlike the Schwarzschild case. This is already 
indicated by the fact that, for $|LQ| \neq 0$, $\lim_{r \rightarrow 0} \epsilon 
= + \infty$ irrespective of the value of $u$. So, no physical particles
with finite energy can access the collision manifold and,
the larger the value of $\epsilon$ they have, the closer to the collision
manifold $r=0$ they can get.

To analyze the repulsive nature of the collision manifold
in more detail 
let us linearize the dynamical system to its first order around the
collision manifold. Thus, we write 
\begin{align*} 
r(s)&=\delta r(s),\\
u(s)&=u_0(s)+\delta u(s),
\end{align*} 
where $u_0(s)$ is a trajectory on the
collision  manifold, so that it satisfies
\begin{equation*} 
u_0'=2 L^2 Q^2+2 u^2.
\end{equation*}
The general solution for this differential equation satisfying $u_0(0)=-\infty$ is
\begin{equation*}
u_0(s)=-|L Q| \cot (2|L Q| s).
\end{equation*}
Where $s\in (0,\frac{\pi }{2 |L Q|})$. The first-order linearized system in the variables  $\delta r$ and
$\delta u$ is
\begin{align*}
\delta r'&= (\delta r) u_0 ,\\
\delta u' &=-3 L^2 M \delta r+4 (\delta u) u_0.
\end{align*}
We can easily solve for $\delta r(s)$:
\begin{equation*}
\delta r(s)=\frac{\delta r_0}{\sqrt{\sin (2 |L Q| s)}},
\end{equation*}
where $\delta r(\frac{\pi }{4 |L Q|})=\delta r_0$. If follows that, no matter
how close to the collision manifold we get (at $s = \frac{\pi}{4 |LQ|}$ where
$\delta r(s)$ is minimum), the trajectory never touches the collision 
manifold and in fact, it separates from it very quickly.
To understand how fast
this happens we need to change to the $\tau$ variable. Given the relation
(\ref{cambiomcgehee}), namely
\begin{equation*}
 \tau'(s)=(\delta r(s))^{3}.
\end{equation*}
As in the Schwarzschild case we will determine a series expansion of
$\delta r(\tau)$ near $\tau = 0$, where $\tau$ is chosen to fulfill $\tau=0$ when $s=\frac{\pi}{4 |L Q|}$. Define $x(s)$ as
\begin{equation*}
x(s)= \frac{1}{\sqrt{\sin (2 |L Q| s)}}
\end{equation*}
so that $(\delta r)(s) = (\delta r_0) x(s)$ and hence
$\frac{dx}{ds} = u_0 x$. In terms of $x$
\begin{equation*}
u_0 = -|L Q| \sqrt{x^4-1},
\end{equation*}
then
\begin{equation*}
\frac{d \tau}{dx} = \frac{d \tau}{ds} \frac{ds}{dx} = \frac{(\delta r_0 
\, x)^{3} }{u_0 x} = 
- \frac{(\delta r_0)^{3}}{|L Q|}
\frac{x^{2}}{\sqrt{ x^4-1}}.
\end{equation*}
Expanding the right-hand side as a series en $x$, integrating and inverting
we find 
\begin{equation*}
x(\tau)=1+(a \tau)^2-\frac{5}{6} (a \tau)^4+\frac{23}{18} (a \tau)^6+ O(\tau)^{8}
\end{equation*}
where $a = \frac{|L Q|}{(\delta r_0)^{3}}$.

It is interesting to note  that inserting 
$ Q = 0 $ in (\ref{colisionRN})
we do not recover the
Schwarzschild case. This is because
the value of the parameter $ \beta $ adapted to 
Reissner-Nordstr\"om is different to that of Schwarzschild. Thus,
in the Schwarzschild subcase of Reissner-Nordstr\"om we have overkilled
the singularity and the
fixed points that previously existed at $ r= 0$, $u = u_{\pm}$
have both collapsed to $ u =0$. This collapse
can be detected directly on the Reissner-Nordstr\"om 
phase space  because the fixed point $\{ u=0, \theta=\theta_0\}$
is no longer hyperbolic when $Q=0$. Another way of seeing  this is
by comparing the shape of the excluded regions of
phase space $ \lbrace r, u \rbrace $ for $ Q = 0 $ with the shape of
the excluded regions in the Schwarzschild phase space. While in the 
latter case the 
excluded regions covered a non-empty interval of $\{ r = 0\} $,
the Reissner-Nordstr\"om regularization is such that the bubble
displayed in Fig. \ref{fig:collision_RN}} , which stays
separated from the collision manifold when $Q \neq 0$,
just touches the line $\{r =0\}$ in the limit $Q =0$.
Thus the whole line of excluded points in the Schwarzschild
regularization has collapsed to a point in the Reissner-Nordstr\"om 
regularization of the Schwarzschild spacetime.

\subsection{The general flow}

\subsubsection{Massless particles}
\begin{figure}[H]
\centering
\centerline{
\begin{tabular}{cccc}
\includegraphics[scale=0.225]{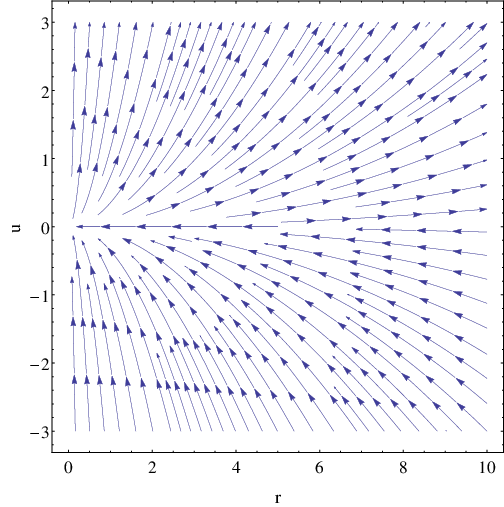}&
\includegraphics[scale=0.225]{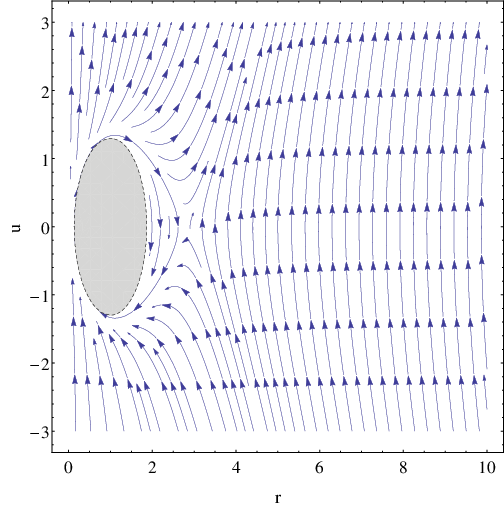}&
\includegraphics[scale=0.225]{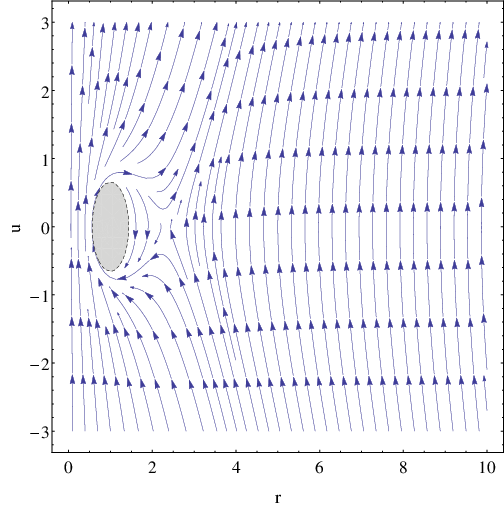} &
\includegraphics[scale=0.225]{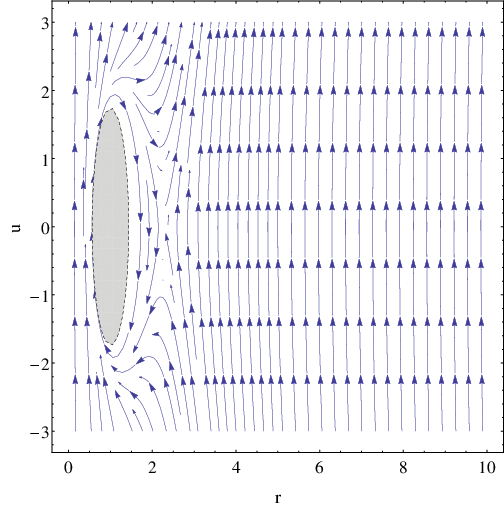}
\end{tabular}}
\vspace*{1em}
\caption{Phase space for massless particles for the values
($L=0$, $Q=0$) , ($L=1.5 M$, $Q=0.5 M$), ($L=1.5 M$, $Q=0.9 M$) and
($L=4 M$, $Q=0.9 M$) from left to right. The dark zone corresponds to the forbidden region.}
  \label{fig:FullphasespaceRNph}
\end{figure}

The reduced dynamical system when $\mu=0$ is
\begin{align*}
r'&=r u\\
u'&=r \left(L^2 r-3 L^2 M\right)+2 \left(L^2 Q^2+u^2\right).
\end{align*}
The phase portrait can be viewed for different values of $L$ and $Q$ in 
Fig. \ref{fig:FullphasespaceRNph}. The fixed points of this system
(assuming $L Q\neq0$) are
\begin{equation*}
(r=R_\pm(M,Q):=\frac{1}{2} \left(3M \pm \sqrt{9 M^2-8 Q^2}\right),u=0)
\end{equation*}
provided $0 < |Q| \leq \sqrt{\frac{9}{8}} M$.  In the strict
inequality case, the two fixed points are hyperbolic, with
the point $(u,r)=(0,R_+(M,Q))$ being a saddle point (with real eigenvalues
of opposite sign) and the point
$(u,r)=(0,R_{-}(M,Q))$ being a center
(purely imaginary eigenvalues).  It is straightforward to check
that the point $(u=0,r=R_{+}(M,Q))$ always lies in the allowed region. 
The point $(u=0,r = R_{-}(M,Q))$ lies in the excluded region as soon
as this region is non-empty, i.e. for $|Q| < M$.

As discussed in Section \ref{excludedRN},
curves encircling the excluded region when $Q \neq 0$ have periodic
properties in the phase space but they are really moving upwards in
the Penrose-Carter diagram changing form the \textit{white hole} patch
to the \textit{black hole} patch as many times as needed. 
Note that with $Q=0$ we recover the Schwarzschild
fixed points but, as  previously noticed in \cref{remarkoverkill},
the phase portrait is nevertheless different
because of the different choice of $\beta$. This is also the case
when  $\mu=1$.

 \subsubsection{Massive particles}

\begin{figure}[H]
\centering
\begin{tabular}{ccc}
\includegraphics[scale=0.225]{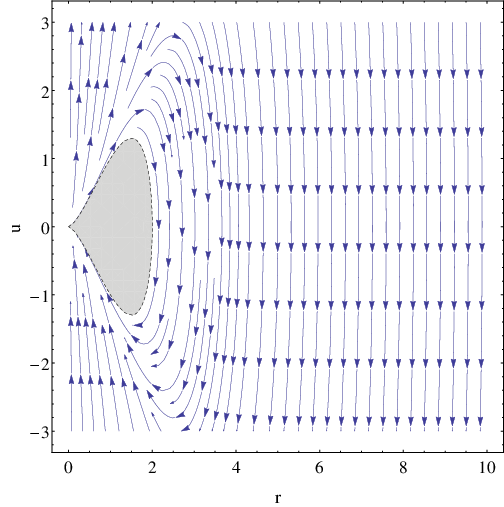}&
\includegraphics[scale=0.225]{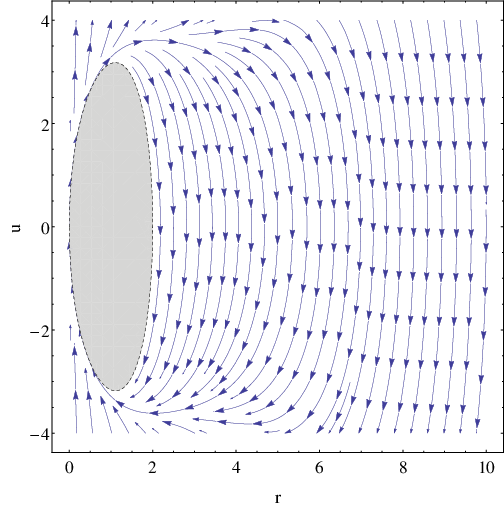}&
\includegraphics[scale=0.225]{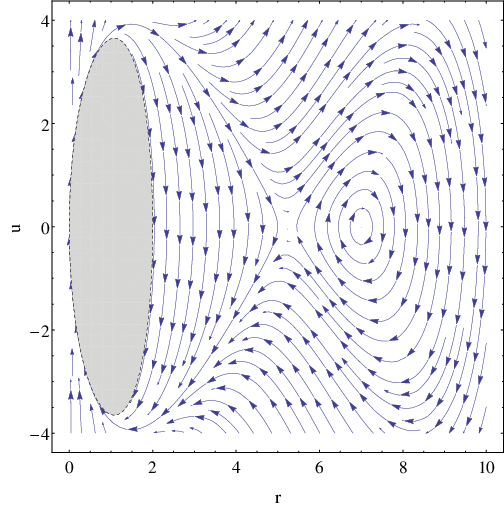}\\
\includegraphics[scale=0.225]{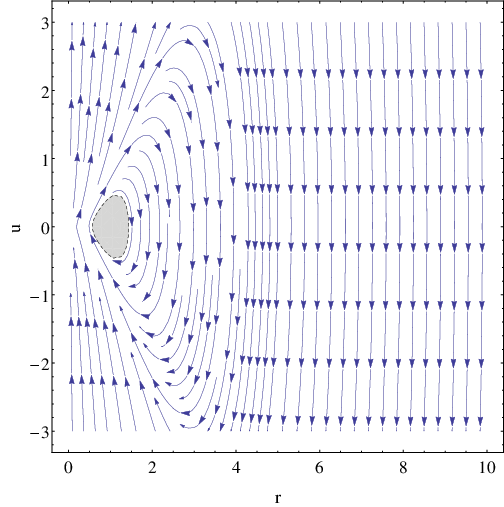}&
 \includegraphics[scale=0.225]{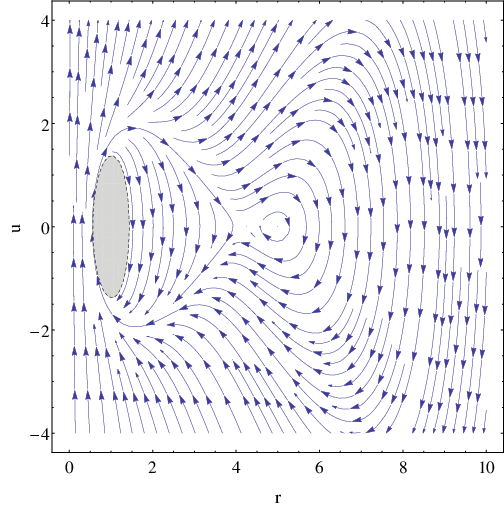}&
\includegraphics[scale=0.225]{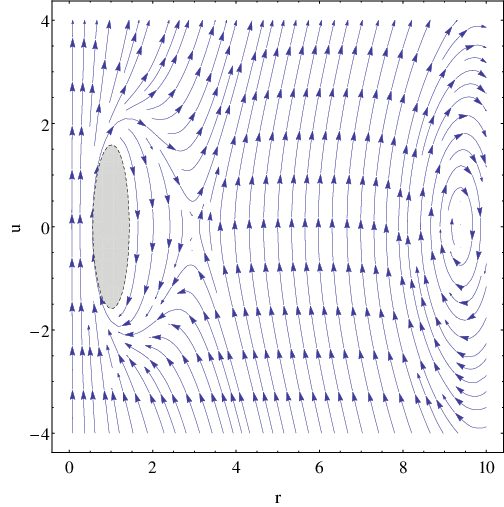}
\end{tabular}
\vspace*{1em}
\caption{Phase space for mass particles with $L=0$, $L=3 M$, $L=3.5 M$ (from left to right) and with $Q=0$ (upper row) y $Q=0.9 M$ (lower row). The dark zone correspond to the forbidden region given by $\epsilon<\frac{\mu}{2}$.}
  \label{fig:FullphasespaceRNp}
\end{figure}
\begin{figure}[H]   
\begin{center}
 \centerline{\includegraphics[scale=0.45]{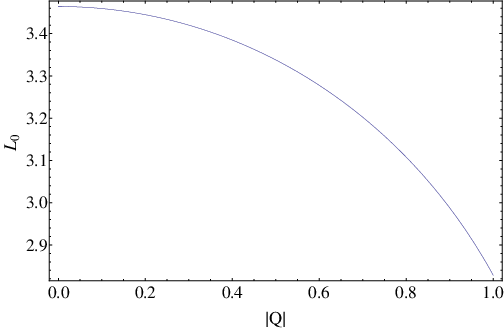}  }
 \end{center}
 \caption{The image shows the variation of the function $L_0$ with respect to the value of $Q$. The plot is given in units where
$M=1$.}
 \label{fig:QL}
\end{figure}

\FloatBarrier
When $\mu=1$ the dynamical system takes the form
\begin{align*}
r'&=r u,\\
u'&=r \left(r \left(L^2-M r+Q^2\right)-3 L^2 M\right)+2 \left(L^2 Q^2+u^2\right),
\end{align*}
with phase spaces displayed for different values of $L$ and $Q$ in Fig.
\ref{fig:FullphasespaceRNp}. The portraits show very clearly
the repulsive nature of the singularity discussed above.
The fixed points lie on the line $u=0$ and are given by the roots of
\begin{equation*}
r^3 M - r^2 \left(L^2+Q^2\right)+3 r L^2 M-2 L^2 Q^2=0.
\end{equation*}
The root structure of this polynomial is not uniform in the parameters
$\{ M,Q,L\}$. Let us concentrate for definiteness in the most interesting case 
$L\neq 0$ and $0<|Q|<M$. It turns out that this equation always has
one real solution
which lies inside the excluded region and corresponds to a hyperbolic critical point that happens to be a center (purely imaginary eigenvalues). 
Moreover, there exists $L_0(M,Q) > 0$ such that, for 
$0< |L| \leq L_0$ this is the only root. For $|L|=L_0$, there is second root
which is double (and hence a  non-hyperbolic fixed point for the dynamical
system). For  $|L| > L_0$ there are two additional hyperbolic points,
both lying outside the excluded region to its right. The one
closer to the excluded region is a saddle and the one with
largest value of $r$ is a center. The function $L_0(M,Q)$
is defined as the only positive and real solution of
\begin{equation*}
 L^6 \left(8 Q^2-9 M^2\right)+ 6 L^4 \left(18 M^4-21  M^2 Q^2+ 4 Q^4\right)+
3 L^2 \left(8 Q^6-3 M^2 Q^4\right)+8 Q^8=0.
\end{equation*}
Existence of a unique positive solution of this equation 
is guaranteed for $0<|Q|<M$.  The function $L_0(Q,M)$ is displayed in Fig. \ref{fig:QL}, note that $L_0(Q=0)=2 \sqrt{3}M$. These points are analogous to the
well-known critical points in the Schwarzschild spacetime and of course they coincide in the limit $Q=0$.

\FloatBarrier

\section{Conclusions}

The main result of this work is a dynamical systems method of analyzing causal
geodesics in stationary and spherically symmetric
Kerr-Schild spacetimes. A remarkable result is that 
the geodesics can be globally described as the motion of a Newtonian particle
in the presence of a radial potential. 
For spacetimes with singularities at the center
we have developed a generalization of the
MacGehee transformation that allows for a regularization of the origin
and hence for a description of the approach to the singularity
in terms of regular variables. In particular, the dynamics
at the collision manifold can be analyzed, which gives us useful
information for the physical trajectories. We have
applied this method to the Schwarzschild and Reissner-Nordstr\"om
spacetimes. Besides the regularized analysis of the
singularity in these spacetimes, we have emphasized the
importance of the presence of excluded regions, which in
effect makes the phase space diagram acquire a non-trivial topology.
This topology and the property that the phase space
portrait is independent of whether we deal with an advanced or a retarded
Kerr-Schild patch allows one to study in the geodesic motion is spacetimes
with complicated global behavior (e.g. the Reissner-Nordstr\"om spacetime)
in terms of a {\it single} two-dimensional phase space portrait.

\section*{Acknowledgements}

 M.M. acknowledges financial support under the projects  FIS2012-30926
(MINECO) and P09-FQM-4496 (Junta de Andaluc\'{\i}a and FEDER funds).\\
P.G. acknowledges the useful comments and help provided by Ester Ramos.

\appendix

\section{Absence of Mcgehee transformation 
decoupling the system in 
\texorpdfstring{$(u,v)$}{(u,v)}} \label{nocoordinates}

In this appendix we prove a no-go Theorem for 
McGehee-type transformations capable of decoupling the dynamic
equations of a point particle on a plane under the influence of a general
radial potential $V(|x|)$.  Specifically, we intend to analyze the 
dynamical system
\begin{align}
\dot{x}&=y \nonumber \\
\dot{y}&= - \nabla V(|x|)=\Lambda(|x|)x.
\label{systemV}
\end{align}
where $\{ x,y \}$ are coordinates on an open subset
of ${\cal N}$ of $\mathbb{C}^2$ and $\nabla =
\partial_{x^1} + i \partial_{x^2}$.
In view of the transformation proposed by
McGehee for the power-law potential (\ref{McGehee}), we define the 
{\it generalized MacGehee  transformation} 
\begin{align}
x &=e^{i \theta } \xi_1(r), \nonumber \\
y &=e^{i \theta } \xi_2(r) (u +i v),  \label{GenMcgehee} \\
d\tau & = \xi_3(r(s)) ds, \nonumber
\end{align}
where $\xi_1,\xi_2, \xi_3 : \mathbb{R}^+ \to \mathbb{R}$ are smooth
and non-zero functions to be determined. Note that $\xi_1(r)$ must be
invertible for this transformation to be well-defined.
The new coordinates
$\{u,v,r,\theta\}$ take values
on $\mathbb{R}$   (for $u,v$), in $\mathbb{R}^{+}$ (for $r$)
and on $S^1$ (for $\theta$).
We note that making $\xi_1$ complex does not define a more general
transformation since it can be reduced to the above one
by redefining the variable $\theta$. Given that we are
replacing the power-law potential $V(|x|) =  |x|^{-\sigma}$ by a general
radial potential, it is reasonable to keep the general
structure of the original McGehee transformation and introduce general
functions of $r$ in the transformation. In this sense, we can consider
(\ref{GenMcgehee}) as the most general Mcgehee transformation in this context.
We prove the following result

\begin{lemma}
Except for potentials which are either power-law
($V(r)= C r^{- \sigma}$)
 or logarithmic ($V(r) = C \mbox{ln} \, r$) 
  there exists no generalized McGehee
transformation capable of decoupling  the system (\ref{systemV})
in the coordinates $(u,v)$.
\end{lemma}
\begin{Proof} 
In the new parameter $s$, the dynamical system is
\begin{align*}
x'&=\xi_3(r) y\\
y'&=\xi_3(r) \Lambda(|x|)x,
\end{align*}
where prime is derivative with respect to $s$. Inserting 
the transformation (\ref{GenMcgehee}) yields 
\begin{align*}
& i \theta' \xi_1 + \frac{d \xi_1}{d r} r' = \xi_2 \xi_3 ( u + i v), \\
& \left ( i \theta' \xi_2 + \frac{d \xi_2}{dr} r' \right )
\left ( u+ i v \right )
+ \xi_2 (u' +i v') = \Lambda(\xi_2) \xi_1 \xi_3.
\end{align*}
Taking real and imaginary parts in the first equation determines
$r'$ and $\theta'$ as
\begin{align}
& r' = \frac{ u \xi_2 \xi_3 }{d\xi_1/dr},  \nonumber \\
& \theta' =\frac{ v \xi_2 \xi_3}{d \xi_1 /dr }. \label{SistemadeMcGehee1}
\end{align}
Substituting into the second equation and separating real
and imaginary parts gives 
\begin{align} 
v'  &= - u  v  \,  \xi_3(r) 
\left( \frac{ d \xi_2 /dr}{d \xi_1/dr}+ \frac{\xi_2}{\xi_1} \right),
\nonumber \\
u' &= \xi_3(r) \left(-\frac{u^2 
d \xi_2/dr}{d \xi_1/dr}+\frac{v^2 \xi_2(r)}{\xi_1(r)} \right) +  \xi_3(r) \frac{\xi_1(r) \Lambda(\xi_1(r))}{\xi_2(r)}. \label{SistemadeMcGehee2}
\end{align}
To uncouple the system in $(u,v)$ it is necessary that
no function of $r$ appears in the equations for $(u', 
v')$. From the 
equation for  $u'$, we need to impose
\begin{align}
\xi_3(r) \left(\frac{d \xi_2/dr}{d \xi_1/dr} \right)&= \beta
\alpha \nonumber  \\
\frac{ \xi_3(r) \xi_2(r)}{\xi_1(r)} &= \alpha, \label{rel1}
\end{align}
where $\alpha, \beta$ are constants with $\alpha\neq 0$.
The second one fixes
$\xi_3$ as
\begin{equation*}
\xi_3(r)=\frac{\alpha \xi_1(r)}{\xi_2(r)},
\end{equation*}
which inserted into the first one gives
\begin{equation*}
\frac{\xi_1(r) d\xi_2/dr}{\xi_2(r) d\xi_1/dr}= \beta
\end{equation*}
This equation can be integrated to obtain:
\begin{equation}
\xi_2(r) = c \, \xi_1(r)^{\beta}. \label{rel2}
\end{equation}
where $c \neq 0$ is a constant. Inserting
these expressions into (\ref{SistemadeMcGehee1})-(\ref{SistemadeMcGehee2}) the dynamical system becomes
\begin{align}
r'  &= \alpha u \frac{\xi_1}{d \xi_1/dr}
\nonumber \\
\theta '  &= \alpha v \nonumber \\
v'  &= - \alpha (1 + \beta)  u v  \label{BestDecoupling} \\
u'  &= \frac{\alpha}{c^2} \xi_1(r)^{2 ( 1- \beta)} \Lambda (\xi_1(r)) 
+ \alpha \left ( v^2 - \beta u^2 \right ). \nonumber
\end{align}
From the expression for $u'$,
the system is uncoupled if and only if 
\begin{equation*}
\xi_1(r)^{2 ( 1 - \beta)}  \Lambda (\xi_1(r)) = \alpha_3
\end{equation*}
for some constant $\alpha_3$, i.e. 
$\Lambda(r)$ is a power-law. Since the potential $V$
is related to $\Lambda$ by
\begin{equation*}
\frac{d V}{dr} = - \Lambda(r) r
\end{equation*}
it follows that the only case for which 
the generalized McGehee transformation decouples the system is
when the potential itself is a power-law or logarithmic 
(recall that an additive
constant is completely irrelevant in the potential $V$).
\end{Proof}



\bibliography{Bibliography}

\end{document}